\documentclass[journal]{IEEEtran}
\ifCLASSINFOpdf
\else
\fi
\usepackage[cmex10]{amsmath}
\usepackage{amsfonts,amssymb}
\newtheorem{theorem}{Theorem}
\newtheorem{definition}{Definition}
\newtheorem{lemma}{Lemma}
\newtheorem{remark}{Remark}
\newtheorem{example}{Example}
\newtheorem{corollary}{Corollary}

\usepackage{graphicx}
\usepackage{subfigure}

\usepackage{booktabs}       
\usepackage{nicefrac}       
\usepackage{xcolor}         

\usepackage{cite}

\begin{document}
\title{Higher-order Common Information} %

\author{Jan {\O}stergaard,~\IEEEmembership{Senior Member,~IEEE}\thanks{J. {\O}stergaard ({jo@es.aau.dk}) is with the 
Department of Electronic Systems, Aalborg University, Aalborg, Denmark. A conference version of this work has been submitted to ITW~\cite{ostergaard:2027a}.}}

%
%

\markboth{Journal of \LaTeX\ Class Files. Manuscript Revised October 2, 2026.}%
{Shell \MakeLowercase{\textit{et al.}}: Bare Demo of IEEEtran.cls for Journals}
%



\maketitle



\begin{abstract}
Shannon mutual information quantifies dependence between two random
variables. We introduce higher-order common information (HCI), a measure of
statistical relevance across $n$ random variables.  The HCI constructs a sequence of stochastically degraded representations. 
At each stage, the representation maximizes the amount of relevant information to the next source under a limited information budget. Due to this limited budget, some source-irrelevant information are potentially  omitted at each stage. The construction can therefore be interpreted as a stochastic relaxation of the intersection of information. 
For finite-alphabet sources, the HCI is lower-bounded by the multivariate
Gács--Körner common information and, for arbitrary sources, it is
upper-bounded by their smallest pairwise mutual information.  We derive closed-form HCI expressions for scalar jointly Gaussian sources and for a binary common-source model.
\end{abstract}


%
\IEEEpeerreviewmaketitle

\section{Introduction}

%
%
%
%

\IEEEPARstart{C}{haracterizing} the information that is \emph{shared} within a collection of $n$ random variables $X_1,\dotsc,X_n$ remains challenging for $n>2$. For $n=2$, Shannon's mutual information provides a measure of pairwise dependence, namely $I(X_1;X_2)$ \cite{Cover:2006}. However, for $n>2$ there is no single canonical extension: different operational questions lead to different notions of ``common information,'' and pairwise summaries can be misleading.
A simple baseline is to quantify redundancy through the smallest pairwise mutual information. For example, for $n$ variables one may define the redundancy $R$ as:
\begin{equation}\label{eq:pairwise}
R \triangleq \min_{i\neq j} I(X_i;X_j).
\end{equation}
While \eqref{eq:pairwise} captures the \emph{amount} of pairwise dependence, it does not determine whether the \emph{same} piece of information is present in all variables. 

For two variables, several inequivalent notions of common information are already classical. 
Wyner's common information \cite{1055346} (and related formulations in, e.g., \cite{witsenhausen,xu:2011}) quantifies the minimum rate of common randomness needed to \emph{simulate} the joint distribution of $(X_1,X_2)$ via a latent variable $W$ that renders them conditionally independent, i.e., $X_1-W-X_2$. Formally,
\begin{align}
C_W(X_1,X_2) &= \inf_{p(w|x_1,x_2)} I(X_1,X_2;W) \notag\\
&\quad \text{s.t.}\quad I(X_1;X_2|W)=0,
\end{align}
which differs from mutual information, e.g., \cite{xu:2011}:
\begin{align}
C_W(X_1,X_2) &= \frac{1}{2}\log_2\!\left(\frac{1+|\rho|}{1-|\rho|}\right),\\
I(X_1;X_2) &= \frac{1}{2}\log_2\!\left(\frac{1}{1-\rho^2}\right).
\end{align}
Extensions and variations of $C_W$ to $n$ variables exists, and in general it holds that $C_W(X_1,X_2) \geq I(X_1;X_2)$ \cite{liu2010common,xu2016lossy,li2016distributed}.

The G\'acs--K\"orner common information \cite{gacs} captures the \emph{deterministic} common part extractable from $X_1$ and $X_2$ without communication. It is defined as the maximum entropy of a random variable $V$ that can be computed as a function of each observation, that is:
\begin{equation}
C_{\mathrm{GK}}(X_1;X_2)\triangleq \max_{f,g:\ f(X_1)=g(X_2)=V} H(V),    
\end{equation}
which can be zero even when $I(X_1;X_2)$ is large. 

For many jointly continuous distributions, entropy-based notions of common randomness can be infinite or degenerate, which motivates alternative characterizations.
Recently, \cite{hanna2025commoninfoDimension} introduced the common information dimension, which replaces rate in bits by information dimension in order to quantify common randomness for continuous random variables in a manner that remains meaningful when entropy-based quantities diverge. 

If one is instead interested in redundancy about a third variable $Y$ (i.e., the information that $X_1,\dotsc,X_n$ share regarding $Y$), then the problem connects to partial information decomposition (PID) and related measures of redundant, unique, and synergistic information \cite{Williams:2010,Harder:2013a,Bertschinger:2014,James:2017,Ince:2017,Gutknecht:2021,ostergaard:2024b}. The BROJA redundancy \cite{Bertschinger:2014} is an  example of a PID redundancy measure defined for discrete alphabets. Extensions to continuous sources are nontrivial and typically rely on restrictive model-based assumptions, such as joint Gaussianity \cite{Barrett:2015,Ince:2017}. 

There also exist multivariate measures that capture global dependencies such as the total correlation \cite{watanabe1960information,han1978nonnegative} and the dual total correlation \cite{han1978nonnegative}, which measures the total amount of statistical dependence among all variables and how much information each variable shares with the rest, respectively.

In this paper, we introduce a notion of higher-order common information (HCI),  $R(X_1,\dotsc, X_n)$, for a
collection of $n$ random variables. Unlike PID
frameworks, the HCI characterizes dependence intrinsic to the collection $\mathbf X=(X_1,\ldots,X_n)$
without introducing an external target variable.
The construction is motivated by the idea of a sequential statistical
"intersection". Starting from $T_1=X_1$, one would ideally retain in
$T_2$ only the information in $T_1$ that is also present in $X_2$.
One would then retain in $T_3$ only the information in $T_2$ that is
also present in $X_3$, and continue in this manner through the complete
source collection. If we momentarily assume that "present in" is interpreted as \emph{deterministic} recoverability, then this sequential procedure corresponds to intersecting the $\sigma$-fields generated by the sources \cite{Kallenberg2021}. For discrete random
variables, this leads to the Gács--Körner common component. Thus, one would recover the multivariate notion $C_{\mathrm{GK}}(\mathbf X)$ of the Gács--Körner common information.
 However, such a deterministic intersection is trivially equal to zero for both noisy discrete sources and for
many continuous sources even though they exhibit strong statistical dependence \cite{gacs}. To be able to capture such statistical dependencies,  our notion of HCI replaces this exact deterministic intersection by a sequential
information-bottleneck construction. In the first stage, set $T_1=X_1$. In the second stage, the next auxiliary is generated from $T_1$
through a stochastic kernel satisfying $T_2-T_1-\mathbf X$. 
The amount of information passed from $T_1$ to $T_2$ is constrained
by:
\begin{equation}
I(T_2;T_1)\leq I(T_1;X_2),
\end{equation}
and $T_2$ is selected to maximize its
relevance to $X_2$:
\begin{equation}
T_2
\in
\arg\sup_{\substack{
S-T_1-\mathbf X\\
I(S;T_1)\leq I(T_1;X_2)}}
I(S;X_2).
\end{equation}
The procedure is then repeated: at stage $\ell$, the representation
$T_{\ell+1}$ is obtained by stochastically degrading $T_\ell$, its
information rate is limited by the relevance of $T_\ell$ to the
current source, say $X_\ell$, and it is chosen to preserve as much of that relevance
as possible.

Our construction of HCI should be understood as a statistical relaxation of
set intersection rather than as a literal intersection of source
components. It does not generally produce a random variable that is
deterministically recoverable from every source. Instead, it measures
how much statistical relevance that survives through a sequence of
locally generated, source-dependent information bottlenecks.
If $T^\star$ is an optimal final stage auxiliary for the source $\mathbf X$ and the HCI is $R(X_1,\dotsc, X_n) = r$, then:
\begin{equation}
I(T^\star;X_j) \geq r,\quad j=1,\dotsc, n.
\end{equation}
Thus, every individual source has at least $r$ bits of
statistical information in common with a representation generated \emph{locally} from
one source. 

For finite-alphabet sources, we show that:
\begin{equation}
R(X_1,\ldots,X_n)
\geq
C_{\mathrm{GK}}(X_1;\ldots;X_n).
\end{equation}
Thus, the HCI preserves every deterministically common component captured
by the Gács--Körner common information, while potentially assigning a
positive value to sources that share no deterministic
common function. For example, for correlated Bernoulli sources, the Gács–Körner common information is zero.  In contrast, the proposed HCI is strictly positive for all finite $n$. Thus, $R(X_1,\dotsc, X_n)$ can reveal shared structure that is not captured by existing measures.  

 To define and characterize $R(X_1,\ldots,X_n)$ for $n\geq3$, we
introduce an iterative information-bottleneck construction based on
nested Markov-chain, information-matching, and target-minimality
constraints. We derive closed-form expressions for jointly Gaussian
sources and for a binary common-source model. In the equicorrelated
cases, these expressions reduce to:
\begin{equation}
R(X_1,\ldots,X_n)
=
-\frac12\log_2\!\left(1-|\rho|^{\,2^{n-1}}\right)
\end{equation}
and
\begin{equation}
R(X_1,\ldots,X_n)
=
1-h_2\!\left(\frac{1-c^{\,2^{n-1}}}{2}\right),
\end{equation}
respectively, where $\rho$ is the Gaussian correlation coefficient
and $c=1-2p$ is the bias of the binary observation channel. For
general sources, we furthermore propose an HCI-inspired numerical approximation
and illustrate it using EEG recordings.

\subsection{Notation}
We use calligraphic letters such as $\mathcal I_i\triangleq \{1,\ldots,n\}\setminus\{i\}$ for sets. Here $\mathcal{I}_i$
is the set of $n-1$ integers not containing $i$. We use boldfaced letters for vectors such as $\mathbf X=(X_1,\dotsc, X_n)$.
We use capital letters such as $X$ and $Y$ for random variables, and $X\perp Y$ is used to indicate that $X$ and $Y$ are mutually independent.
We denote by $X \sim \mathrm{Bern}(p)$ a Bernoulli random variable with
$\mathbb{P}(X=1)=p, \mathbb{P}(X=0)=1-p, p \in [0,1]$.
The binary entropy function is defined as $
h_2(p) \triangleq -p \log_2 p - (1-p)\log_2(1-p)$,
with the convention $0\log_2 0 = 0$. 
For $a,b \in [0,1]$, the binary convolution is $a \star b \triangleq a(1-b) + (1-a)b.$ 
For $a,b \in \{0,1\}$, the modulo-2 addition is $
a \oplus b \triangleq (a+b) \bmod 2.$
Let $(a_1,\dots,a_n) \in \mathbb{R}^n$. The order statistics are the sorted values $(a_{(1)},\dots,a_{(n)})$ such that $a_{(1)} \le \cdots \le a_{(n)}.$
Let $\pi^{(i)}$ be a permutation of $\mathcal I_i$. 
With this notation, $X_j = X_{\pi^{(i)}(\ell)}$ if $j$ is the $\ell$th element of $\mathcal I_i$, that is $j=\pi^{(i)}(\ell)$ under the permutation $\pi^{(i)}$ of $\mathcal I_i$.

\subsection{Paper organization}
In Section \ref{sec:hci}, we present our new notion called HCI. In Section \ref{sec:gaussian}, we consider Gaussian sources, and in Section~\ref{sec:discrete} we consider some discrete sources. In Section \ref{sec:general_sources} we demonstrate how to apply our results on real data. Longer proofs of theorems are in the appendix.

\section{Higher-order common information}\label{sec:hci}
In this section, we formally introduce our notion of HCI, we provide upper and lower bounds, and finally present some key properties of the HCI. 

\subsection{Definition of HCI}
For an arbitrarily distributed random variable $X$, define:
\begin{equation}
\begin{aligned}
\mathsf M(X)\triangleq
\bigl\{S:\;&
\text{for every measurable }G=g(S),\\
&X-G-S\ \Longrightarrow\
S=f(G)\ \text{a.s.}\\
&\text{for some measurable }f
\bigr\}.
\end{aligned}
\label{eq:MX}
\end{equation}

\begin{example}
Let $E$ and $X$ be arbitrarily distributed and mutually independent. Set $T=X$ and define $T'=(T,E)$.
Then $T'$ contains a separately accessible source-irrelevant
component. Indeed, with $G=T$, we have $G=g(T'), \mathbf X-G-T'$,
but $T'$ is not a measurable function of $G$, since $E$ is
 independent of $G$. Hence $T'\notin\mathsf M( X)$.
On the other hand, $T=X$ belongs to
$\mathsf M( X)$, since $T$ is equal to the source $X$.
\end{example}

\begin{example}
Suppose that $ X $ is Gaussian, $T=X$, and  $\widetilde T=aT+E, a \ne 0$,
where $E$ is Gaussian and independent of $ X$.
Then $\widetilde T$ is target-minimal with respect to $ X$.
Although $E$ is target-independent, it is not available as a separate
component of $\widetilde T$, since it is mixed with $T$.
Moreover, the conditional Gaussian distribution
$P_{ X|\widetilde T=t}$ has a mean that is a nonzero linear
function of $t\neq 0$. It follows that distinct values of $\widetilde T$
results in distinct conditional distributions of $ X$. Therefore,
any statistic $G=g(\widetilde T)$ satisfying $ X-G-\widetilde T$
must determine $\widetilde T$ almost surely, proving that $\widetilde T\in\mathsf M( X)$.
\end{example}

\begin{definition}[Higher-order common information]
\label{def:HCI}
Let $\mathbf X=(X_1,\ldots,X_n)$ be a random vector whose components take values in standard Borel
spaces. Fix a reference index $i\in\{1,\ldots,n\}$, let 
$\pi^{(i)}$ be a permutation of $\mathcal I_i\triangleq\{1,\ldots,n\}\setminus\{i\}$,
and set $T_1^{(i)}=X_i$.
For $\ell=1,\ldots,n-2$, given $T_\ell^{(i)}$, define:
\begin{align}\notag
&\mathcal F_\ell(T_\ell^{(i)},\pi^{(i)},\mathbf X) \\ \label{eq:stage_feasible_set}
&\triangleq
\left\{
S:
\begin{array}{ll}
\text{(MC)\ } & S-T_\ell^{(i)}-\mathbf X,\\[0.2em]
\text{(IM)\ } & I(S;T_\ell^{(i)})
\leq
I(T_\ell^{(i)};X_{\pi^{(i)}(\ell)}),\\[0.2em]
\text{(TM)\ } & S\in\mathsf M(X_{\pi^{(i)}(\ell)}).
\end{array}
\right\}.
\end{align}
The three conditions in \eqref{eq:stage_feasible_set} are called the
Markov-chain (MC), information-matching (IM), and target-minimality (TM) constraints. 
Define the optimal stage-wise source relevance by:
\begin{equation}\label{eq:stage_optimal_value}
F_\ell(T_\ell^{(i)},\pi^{(i)},\mathbf X)
\triangleq
\sup_{S\in\mathcal F_\ell(T_\ell^{(i)},\pi^{(i)},\mathbf X)}
I(S;X_{\pi^{(i)}(\ell)}),
\end{equation}
and the  optimizer set by:
\begin{equation}
\begin{aligned}
\mathcal A_\ell
\bigl(T_\ell^{(i)},\pi^{(i)},\mathbf X\bigr)
&\triangleq
\Bigl\{S\in{}
\mathcal F_\ell
\bigl(T_\ell^{(i)},\pi^{(i)},\mathbf X\bigr):\\[-0.2em]
&I\bigl(S;X_{\pi^{(i)}(\ell)}\bigr)
=
F_\ell
\bigl(T_\ell^{(i)},\pi^{(i)},\mathbf X\bigr)
\Bigr\}.
\end{aligned}
\label{eq:stage_optimizer_set}
\end{equation}

The terminal set is defined as:
\begin{equation}
\mathcal T_i^{\pi^{(i)}}(\mathbf X)
\triangleq
\left\{
T_{n-1}^{(i)}:
\begin{array}{l}
T_1^{(i)}=X_i,\\[0.2em]
T_{\ell+1}^{(i)}
\in
\mathcal A_\ell(T_\ell^{(i)},\pi^{(i)},\mathbf X),\\[0.2em]
\ell=1,\ldots,n-2
\end{array}
\right\}.
\end{equation}
We can now finally define the HCI by:
\begin{equation}\label{eq:Rn}
R(X_1,\ldots,X_n)
\triangleq
\min_i
\max_{\pi^{(i)}}
\sup_{T\in\mathcal T_i^{\pi^{(i)}}(\mathbf X)}
\min_j I(T;X_j).
\end{equation}
\end{definition}

\begin{remark}
Note that this HCI formulation has only a solution when all stage-wise suprema are attained, so that the corresponding optimizer sets are nonempty. 
Throughout this paper, $R(X_1,\dotsc, X_n)$ is therefore defined only for sources for which the stage-wise optimizer sets are nonempty. This condition is verified for the Gaussian and binary source models considered below. An extension to general standard Borel sources, where the suprema are not attained could, for example, be obtained by replacing the supremum by an $\varepsilon$-optimal maximizer. Such extensions are left for future work.
\end{remark}

We will now discuss the constraints in the HCI definition in more details. 
The nested MC and IM constraints in \eqref{eq:stage_feasible_set}
play complementary roles. The MC constraints:
\begin{equation}
T_{\ell+1}^{(i)}-T_\ell^{(i)}-\mathbf X,
\quad
T_1^{(i)}=X_i,
\end{equation}
ensure that every auxiliary is generated by successively processing a
single reference source $X_i$. In particular, the construction has never access to multiple source variables and therefore cannot
introduce information that is available only through their joint observation such as synergistic information. Thus, the Markov chain $T_{\ell}^{(i)} - X_i - \mathbf X$ holds for all $i$ and $\ell$.

The IM constraints:
\begin{equation}
I(T_{\ell+1}^{(i)};T_\ell^{(i)})
\leq
I\!\left(T_\ell^{(i)};
X_{\pi^{(i)}(\ell)}\right)
\end{equation}
successively restrict the amount of information that can be passed from
one auxiliary to the next. These constraints prevent the construction
from carrying the reference variable unchanged through the entire
sequence. Indeed, if the IM constraints were omitted, the identity
chain:
\begin{equation}
T_1^{(i)}=T_2^{(i)}=\cdots=T_{n-1}^{(i)}=X_i
\end{equation}
would be feasible. For a fixed reference $X_i$, the resulting terminal
optimization would then satisfy:
\begin{equation}
\sup_{T:\,T-X_i-\mathbf X}\min_j I(T;X_j)
=
\min_j I(X_i;X_j).
\end{equation}
Thus, the sequential IM constraints prevent the HCI from
degenerating to the smallest pairwise mutual information of the source variables. 

The interplay between the MC and IM constraints and the stage-wise relevance maximization admits an informal interpretation as a stochastic relaxation of
information intersection. At each stage, the new auxiliary is generated from the preceding representation, while the IM constraint
limits the information passed on. Subject to this budget, the optimization selects an auxiliary that preserves as much source-relevant information as possible and thereby potentially reduces the amount of source-irrelevant information. Note that unlike the case of exact intersection of deterministic information, our construction will not remove \emph{all} source-irrelevant information. The MC and IM constraints restrict how information is propagated but do not, by themselves, enforce a minimal representation.
For example, suppose that $T$ is feasible and define $T'=(T,E)$, where $E\perp(T,\mathbf X)$. Appending $E$ does not change the
statistical relationship between the auxiliary and the sources:
\begin{equation}
I(T';X_j)=I(T;X_j),
\quad j=1,\ldots,n,
\end{equation}
and $I(T';\mathbf X)=I(T;\mathbf X)$. 
Although such private randomness does not change the current stage-wise supremum, it may change the feasible set
at subsequent stages, because a later auxiliary can be generated from the complete representation $T'=(T,E)$. To remove deterministically
separable components that do not contribute to the statistical relationship with the current target, we impose the target-minimality
(TM) constraint. In the case $n=3$, there is only one auxiliary stage, and appending independent private randomness to the
terminal auxiliary does not change its mutual information with any source. For $n\geq4$, however, intermediate auxiliaries are used to
generate subsequent representations, so redundant components can propagate through the sequence and complicate both the optimization and
the characterization of the optimal auxiliary chain. We therefore introduce the TM constraint to exclude such nonessential representations.

We allow both deterministic and stochastic mappings. For many sources, stochastic mappings are superior. 
The following example shows that deterministic mappings do not necessarily saturate the IM constraint, whereas a stochastic mapping can attain
equality and yield a strictly larger HCI value.

\begin{example}
\label{ex:erasure_HCI}
Let $n=3$, and let $X_1$ take values in $\{0,1,2\}$ with $\mathbb{P}_{X_1}(0)=\frac12$ and $\mathbb{P}_{X_1}(1)=\mathbb{P}_{X_1}(2)=\frac14$.
Moreover, define
\begin{equation}
\label{eq:X2_erasure}
X_2=X_3=
\begin{cases}
X_1,&B=1,\\
\mathtt e,&B=0,
\end{cases}
\end{equation}
where $B\sim\operatorname{Bern}(\alpha)$, with
$\alpha=3/5$, is independent of $X_1$, and
$\mathtt e\notin\{0,1,2\}$ is an erasure symbol.

Since $H(X_1)=\frac32$, and $X_2$ and $X_3$ are identical erasure observations of $X_1$, we have that:
\begin{equation}
I(X_1;X_2)
=
I(X_1;X_3)
=
\alpha H(X_1)
=
\frac9{10}.
\label{eq:erasure_pairwise_mi}
\end{equation}

Fix $X_1$ as the reference and process $X_2$ first. A feasible
auxiliary $S$ must satisfy $S\in\mathsf M(X_2)$ and:
\begin{equation}
S-X_1-(X_1,X_2,X_3),
\quad
I(S;X_1)\leq \frac9{10}.
\label{eq:erasure_IM_budget}
\end{equation}
For every $S-X_1-X_2$, we have that:
\begin{align}
H(S|X_2)
&=(1-\alpha)H(S)+\alpha H(S|X_1),
\end{align}
and therefore:
\begin{equation}
I(S;X_2)
=
\alpha I(S;X_1).
\label{eq:erasure_scaling_HCI}
\end{equation}
Because $X_2=X_3$, we also have that $I(S;X_3)=\alpha I(S;X_1)$.

We first restrict $S$ to be a deterministic function of $X_1$.
Every such function corresponds, up to relabeling of its outputs, to a
partition of $\{0,1,2\}$. The possible partitions are listed in
Table~\ref{tab:deterministic_partitions}.
\begin{table}[t]
\centering
\caption{Deterministic partitions of the alphabet of $X_1$.}
\label{tab:deterministic_partitions}
\begin{tabular}{c c c}
\hline
Partition & Distribution of $S$ & $I(S;X_1)=H(S)$\\
\hline
$\{0,1,2\}$
& $1$
& $0$\\
$\{0\}\mid\{1,2\}$
& $(\frac12,\frac12)$
& $1$\\
$\{1\}\mid\{0,2\}$
& $(\frac14,\frac34)$
& $h_2(\frac14)$\\
$\{2\}\mid\{0,1\}$
& $(\frac14,\frac34)$
& $h_2(\frac14)$\\
$\{0\}\mid\{1\}\mid\{2\}$
& $(\frac12,\frac14,\frac14)$
& $\frac32$\\
\hline
\end{tabular}
\end{table}
None of the deterministic partitions attains the upper limit $I(S;X_1)=\frac9{10}$ 
because:
\begin{equation}
h_2\!\left(\frac14\right)
\approx0.8113
<
\frac9{10}.
\end{equation}
The largest feasible  value is therefore
$h_2(1/4)$, attained, for example, by the partition $\{1\}\mid\{0,2\}$, corresponding to $S_{\mathrm{det}} = 1$ if $X_1=1$ and $S_{\mathrm{det}} = 0$ if $X_1\in \{0,2\}$. 
The two values of $S_{\mathrm{det}}$ induce distinct conditional
distributions of $X_2$, so $S_{\mathrm{det}}\in\mathsf M(X_2)$.
Using \eqref{eq:erasure_scaling_HCI}, its current-source relevance is:
\begin{equation}
I(S_{\mathrm{det}};X_2)
=
\alpha h_2\!\left(\frac14\right)
\approx0.4868\ \text{bits}.
\label{eq:deterministic_reference_X1}
\end{equation}
Hence this is the largest stage-wise value obtainable by a
deterministic auxiliary when $X_1$ is the reference.

We now allow stochastic mappings. Let
$C\sim\operatorname{Bern}(\lambda)$ be independent of
$(X_1,X_2,X_3)$, and define
\begin{equation}
S^\star=
\begin{cases}
X_1,&C=1,\\
\mathtt e_S,&C=0,
\end{cases}
\label{eq:stochastic_erasure_auxiliary}
\end{equation}
where $\mathtt e_S$ is a new erasure symbol. Then $I(S^\star;X_1)=\lambda H(X_1)$.
Choosing $\lambda=\alpha=\frac35$ gives:
\begin{equation}
I(S^\star;X_1)
=
\alpha H(X_1)
=
\frac9{10}
=
I(X_1;X_2).
\label{eq:stochastic_IM_saturation}
\end{equation}
Thus, this choice of stochastic auxiliary attains equality in the IM constraint.
Moreover, by \eqref{eq:erasure_scaling_HCI}, we have that:
\begin{align}
I(S^\star;X_2)
&=I(S^\star;X_3)\\
&=\alpha I(S^\star;X_1)\\
&=\alpha^2H(X_1)\\
&=\frac{27}{50}\\
&=0.54\ \text{bits}.
\label{eq:stochastic_achievability}
\end{align}
This auxiliary is stage-wise optimal. Indeed, every feasible $S$
satisfies:
\begin{equation}
I(S;X_2)
=
\alpha I(S;X_1)
\leq
\alpha I(X_1;X_2)
=
\alpha^2H(X_1),
\end{equation}
and $S^\star$ attains this upper bound.

It remains to verify target minimality. The conditional distributions $P_{X_2|S^\star=s}$ are distinct for the four possible values
$s\in\{0,1,2,\mathtt e_S\}$. Hence no two values of $S^\star$ can be merged without changing its statistical relationship
with $X_2$, which proves that $S^\star\in\mathsf M(X_2)$.
Therefore, for reference $X_1$, the stage-wise optimizer $S^\star$
has terminal value:
\begin{equation}
\min_{j\in\{1,2,3\}}I(S^\star;X_j)
=
\frac{27}{50}.
\label{eq:reference_X1_value}
\end{equation}
The same value is obtained when $X_3$ is processed instead of $X_2$,
because $X_2=X_3$.

Now choose $X_2$ as the reference and process $X_3$ first. Since
$X_2=X_3$, the identity auxiliary $S=X_2$ is feasible, target-minimal, and stage-wise optimal. Its terminal value
is:
\begin{equation}
\min_j I(X_2;X_j)
=
I(X_2;X_1)
=
\frac9{10}.
\end{equation}
Thus, after maximizing over the two possible orderings, the
fixed-reference value for $X_2$ is $9/10$. By symmetry, the same
holds for reference $X_3$.
Finally, minimizing over the reference variables gives:
\begin{align}
R(X_1,X_2,X_3) = 0.54\ \text{bits},
\end{align}
which is strictly greater than $0.4868$\ bits obtained when the auxiliary mappings are restricted to be deterministic.

For comparison, we note that $I(X_1;X_2)=I(X_1;X_3)=0.9\ \text{bits}$, while $I(X_2;X_3) = H(X_2) \approx1.871$ bits.
Thus, all pairwise mutual informations are strictly larger than the HCI.
\hfill$\triangle$
\end{example}

\subsection{Upper Bound to the HCI}\label{sec:ub}
In this section, we introduce an upper bound to the HCI by removing some of the constraints. 
Specifically, 
we define the following relaxed optimization problem by retaining the
MC and IM constraints from \eqref{eq:stage_feasible_set}, while removing the TM and stage-wise optimality
requirements:
\begin{align}\notag
&R_{\mathrm{ub}}(X_1,\ldots,X_n) \\ \label{eq:Rub}
&\triangleq
\min_{1\leq i\leq n}
\max_{\pi^{(i)}} \!\!\!\!\!\!\!\!\!
\sup_{\substack{
T_1^{(i)}=X_i,\\
T_{\ell+1}^{(i)}-T_\ell^{(i)}-\mathbf X,\\
I(T_{\ell+1}^{(i)};T_\ell^{(i)})
\leq
I\!\left(T_\ell^{(i)};
X_{\pi^{(i)}(\ell)}\right),\\
\ell=1,\ldots,n-2
}}
\!\!\!\!\!\!\!\!\!
\min_{1\leq j\leq n}
I(T_{n-1}^{(i)};X_j),
\end{align}
where the supremum is taken over the auxiliary sequence
$(T_2^{(i)},\ldots,T_{n-1}^{(i)})$.

The following corollary follows directly from the fact that every auxiliary chain admissible for the HCI is also admissible for the
relaxed optimization problem in \eqref{eq:Rub} which defines $R_{\mathrm{ub}}$.
\begin{corollary}
    For any source $\mathbf X$, we have that:
    \begin{equation}\label{eq:Rub1}
        R_{\mathrm{ub}}(\mathbf X) \geq R(\mathbf X).
    \end{equation}
    \hfill$\triangle$
\end{corollary}
Thus, an auxiliary may keep information that is irrelevant to the current source but relevant to subsequent sources, provided that doing so improves the supremum over the terminal stage auxiliary. 
We provide an example of such a case below, where we further show that the upper bound in \eqref{eq:Rub1} can be strict. 
\begin{example}
\label{ex:triangle_upper_bound}
Let $A,B,C$ be mutually independent fair bits and define $X_1=(A,B), X_2=(A,C),$ and $X_3=(B,C)$.
Each pair of sources shares exactly one bit: 
\begin{equation}
I(X_1;X_2)
=
I(X_1;X_3)
=
I(X_2;X_3)
=
1.
\end{equation}
However, the three sources have no deterministic component in common, which implies that $C_{\mathrm{GK}}(\mathbf X) = 0$. 

We first compute the HCI. Fix $X_1$ as the reference and suppose that
$X_2$ is processed first. A feasible auxiliary $S$ must satisfy:
\begin{equation}
S-X_1-(X_1,X_2,X_3),
\quad
I(S;X_1)\leq I(X_1;X_2)=1.
\label{eq:triangle_SW_feasible}
\end{equation}
Since $C\perp(A,B,S)$, $I(S;X_2)=I(S;A)$.
Therefore, $I(S;X_2)\leq H(A)=1$,
and the upper bound  is attained by $S=A$. Hence the maximum
stage-wise relevance is one bit. Now let $S^\star$ be any stage-wise optimizer. Then $I(S^\star;A)=1$,
which implies that $H(A|S^\star)=0$.
Moreover, we have that:
\begin{align}
I(S^\star;X_1)
&=I(S^\star;A,B)\\
&=I(S^\star;A)+I(S^\star;B|A)\\
&=1+I(S^\star;B|A).
\end{align}
The IM constraint $I(S^\star;X_1)\leq1$ therefore implies $I(S^\star;B|A)=0$.
Thus, $S^\star-A-B$.
Since $A\perp B$, it follows that $S^\star\perp B$. In addition,
$C\perp(A,B,S^\star)$, and hence $I(S^\star;X_3) = I(S^\star;B,C) = 0$.
Therefore, every stage-wise optimizer has terminal value:
\begin{equation}
\min_{j\in\{1,2,3\}}I(S^\star;X_j)=0.
\end{equation}

The same argument applies when $X_3$ is processed first and, by
symmetry, for every choice of reference variable and ordering.
It therefore follows that: 
\begin{equation}
R(X_1,X_2,X_3)=C_{GK}(\mathbf X) = 0.
\label{eq:triangle_HCI_zero}
\end{equation}

Let us now compute the relaxed upper bound $R_{\mathrm{ub}}$. Again fix
$X_1$ as the reference. Every relaxed feasible auxiliary $T$
satisfies:
\begin{equation}
T-X_1-(X_1,X_2,X_3),
\quad
I(T;X_1)\leq1.
\end{equation}
Because $C\perp(A,B,T)$, we have that:
\begin{equation}
I(T;X_2)=I(T;A),
\quad
I(T;X_3)=I(T;B).
\end{equation}
Using $A\perp B$, we obtain:
\begin{align}
I(T;X_2)+I(T;X_3)
&=I(T;A)+I(T;B)\\
&=I(T;A,B)-I(A;B|T)\\
&\leq I(T;A,B)\\
&=I(T;X_1)\\
&\leq1.
\end{align}
It follows that:
\begin{equation}
\min_j I(T;X_j)\leq\frac12.
\label{eq:triangle_upper_converse}
\end{equation}
To attain the upper bound in \eqref{eq:triangle_upper_converse}, let
$Q\sim\operatorname{Bern}(\tfrac12)$ be independent of
$(A,B,C)$, and define:
\begin{equation}
T=(Q,Z),
\quad
Z=
\begin{cases}
A,&Q=0,\\
B,&Q=1.
\end{cases}
\end{equation}
This auxiliary is generated locally from $X_1=(A,B)$, and hence:
\begin{equation}
T-X_1-(X_1,X_2,X_3).
\end{equation}
Moreover, $I(T;X_1)=I(T;A,B)=1$, so the IM constraint is satisfied with equality. Furthermore:
\begin{align}
I(T;X_2)
&=I(T;A)=\frac12,\\
I(T;X_3)
&=I(T;B)=\frac12.
\end{align}
Since $I(T;X_1)=1$, the resulting terminal value is $\min_j I(T;X_j)=\frac12$.
Together with \eqref{eq:triangle_upper_converse}, this proves that for this reference the relaxed value is $1/2$. By symmetry, the same value is
obtained for every reference and ordering. Therefore:
\begin{equation}
R_{\mathrm{ub}}(X_1,X_2,X_3)=\frac12,
\label{eq:triangle_Rub}
\end{equation}
and we have thus obtained:
\begin{align}
R(X_1,X_2,X_3)
&=
0 \\
&<
\frac12 =
R_{\mathrm{ub}}(X_1,X_2,X_3) \\
&<
1
=
\min_{i\neq j}I(X_i;X_j).
\end{align}
The relaxed upper bound balances the $A$- and $B$-components in a
single randomized representation, even though they are relevant to
different sources. On the other hand, stage-wise optimality prevents such a tradeoff:
processing $X_2$ first requires retaining $A$ maximally, which
removes the $B$-information which is irrelevant for $X_2$ but relevant for $X_3$. This example therefore
illustrates both the strictness of $R_{\mathrm{ub}}$ and the role of
stage-wise optimality in the HCI definition.
\hfill$\triangle$
\end{example}

\subsection{Lower Bound to the HCI}
The HCI is trivially lower bounded by zero, since the mutual information is non-negative. For discrete sources with finite alphabets, we can improve the lower bound by connecting the HCI to the Gács--Körner common information. 
\begin{theorem}[Gács--Körner lower bound]
\label{thm:GK_lower_bound_HCI}
Let $X_1,\ldots,X_n$ be finite-alphabet random variables. Then
\begin{equation}
R(X_1,\ldots,X_n)
\geq
C_{\mathrm{GK}}(X_1;\ldots;X_n),
\label{eq:GK_lower_bound_HCI}
\end{equation}
where:
\begin{equation}
C_{\mathrm{GK}}(X_1;\ldots;X_n)
\triangleq
\max_{\substack{V:\\
H(V|X_j)=0,\;j=1,\ldots,n}}
H(V).
\end{equation}
\end{theorem}

\begin{IEEEproof}
Let $V$ be any deterministic common component of the sources such that:
\begin{equation}
H(V|X_j)=0,
\quad j=1,\ldots,n.
\label{eq:V_common_function}
\end{equation}
We first prove that if the current representation contains $V$, then
a stage-wise optimizer can be chosen to contain $V$ as well.

Fix a stage with current representation $T$ and current target $X$,
and suppose that:
\begin{equation}
H(V|T)=H(V|X)=0.
\label{eq:V_in_T_X}
\end{equation}
The available IM budget is:
\begin{align}
\Delta
&\triangleq I(T;X)\\
&=I(V,T;X)\\
&=H(V)+I(T;X|V).
\end{align}
We define $\delta\triangleq I(T;X|V)$, 
so that:
\begin{equation}
\Delta=H(V)+\delta.
\label{eq:stage_budget_V}
\end{equation}

Let $S$ be any auxiliary satisfying:
\begin{equation}
S-T-\mathbf X,
\quad
I(S;T)\leq \Delta.
\label{eq:S_stage_feasible}
\end{equation}
To simplify the notation, we define:
\begin{equation}
h\triangleq H(V|S),\quad
r\triangleq I(S;T|V),\quad
a\triangleq I(S;X|V).
\end{equation}
Since $V$ is a deterministic function of both $T$ and $X$, it follows that:
\begin{align}
I(S;T)&=H(V)-h+r,\label{eq:S_rate_decomp}\\
I(S;X)&=H(V)-h+a.\label{eq:S_relevance_decomp}
\end{align}
Moreover, $S-T-X$ remains a Markov chain conditionally on $V$, and
hence the data processing inequality yields:
\begin{equation}
a\leq r.
\label{eq:a_less_r}
\end{equation}
The feasibility condition and \eqref{eq:S_rate_decomp} imply $H(V)-h+r\leq H(V)+\delta$, which can equivalently be written as:
\begin{equation}
h\geq r-\delta.
\label{eq:h_rate_gap}
\end{equation}

We now construct a feasible auxiliary that explicitly retains $V$.
Conditionally on $V$, independently erase $S$ with retention
probability $\lambda$, and denote the result by $W$. More precisely,
let $E\sim\operatorname{Bern}(\lambda)$ be independent of
$(S,T,\mathbf X,V)$, and set:
\begin{equation}
W=
\begin{cases}
S,&E=1,\\
\mathtt e,&E=0.
\end{cases}
\end{equation}
Moreover, define $S'=(V,W)$. Then $S'-T-\mathbf X$, and we have that:
\begin{align}
I(S';T)
&=H(V)+\lambda r,\label{eq:Sprime_rate}\\
I(S';X)
&=H(V)+\lambda a.\label{eq:Sprime_relevance}
\end{align}

We now choose $\lambda$ such that:
\begin{equation}
\lambda
\triangleq
\begin{cases}
1,&r\leq\delta,\\[0.2em]
\delta/r,&r>\delta.
\end{cases}
\label{eq:lambda_choice}
\end{equation}
It follows from \eqref{eq:Sprime_rate} that:
\begin{equation}
I(S';T)\leq H(V)+\delta=\Delta,
\end{equation}
so $S'$ satisfies the IM constraint.

We next compare the relevance of $S'$ and $S$. If $r\leq\delta$,
then $\lambda=1$, and:
\begin{equation}
I(S';X)=H(V)+a
\geq H(V)-h+a
=I(S;X).
\end{equation}
If $r>\delta$, then $\lambda=\delta/r$. From
\eqref{eq:a_less_r} and \eqref{eq:h_rate_gap}, it follows that:
\begin{equation}
a\left(1-\frac{\delta}{r}\right)
\leq r-\delta
\leq h.
\end{equation}
Therefore, we finally obtain:
\begin{align}
I(S';X)
&=H(V)+\frac{\delta}{r}a\\
&\geq H(V)-h+a\\
&=I(S;X).
\end{align}
Thus, every feasible $S$ can be replaced by a feasible $S'$ that
contains $V$ and has at least as much relevance to the current target.

It remains to enforce target minimality. Because the alphabets are
finite, let $G=g(S')$ be the minimal sufficient statistic of $S'$
for $X$, obtained by grouping values $s'$ that induce the same
conditional distribution $P_{X|S'=s'}$. Then, it follows that:
\begin{equation}
X-G-S',
\quad
I(G;X)=I(S';X),
\end{equation}
and $G\in\mathsf M(X)$. Moreover, since $G$ is a deterministic function of
$S'$:
\begin{equation}
I(G;T)\leq I(S';T)\leq\Delta,
\end{equation}
which implies that $G$ remains feasible.

Since $V$ is a deterministic function of both $X$ and $S'$ and 
 $X-G-S'$, we obtain that:
\begin{equation}
H(V|G)=0.
\label{eq:V_preserved_by_G}
\end{equation}
Hence the target-minimal auxiliary $G$ retains $V$.

We have so far proved that whenever the current
representation $T$ contains $V$, every subsequent feasible auxiliary can be
replaced by a target-minimal feasible auxiliary that also contains
$V$ and has no smaller current-target relevance. Thus, a
stage-wise optimizer can always be selected such that it retains $V$.
By applying this result inductively for each stage, we can always find a feasible and optimal terminal auxiliary $T_{n-1}^{(i)}$
such that $H(V|T_{n-1}^{(i)})=0$.
Since $V$ is also a deterministic function of every $X_j$, we have that:
\begin{align}
I(T_{n-1}^{(i)};X_j)
&\geq I(V;X_j)\\
&=H(V),
\quad j=1,\ldots,n.
\end{align}
This holds for every reference index and every
ordering. Hence, we obtain:
\begin{equation}
R(X_1,\ldots,X_n)\geq H(V).
\end{equation}
Finally, maximizing over all deterministic common components $V$ proves:
\begin{equation}
R(X_1,\ldots,X_n)
\geq
C_{\mathrm{GK}}(X_1;\ldots;X_n).
\end{equation}
\end{IEEEproof}

\subsection{Properties of the HCI}
In this section we establish key properties of the HCI. We begin by the following lemma, which proves that when the supremum is attained in \eqref{eq:Rn}, then the IM constraint must be saturated. Thus, when restricting to sources where the supremum is attainable, one can in principle replace in the IM inequality constraint by an equality constraint without loss of optimality.\footnote{Equality in the IM constraint might not always be possible by deterministic functions, which is why we prefer to keep the inequality constraint.} 
\begin{lemma}
\label{lem:IM_saturation}
Let $S,T,X$ be random variables satisfying $0<I(T;X)<\infty$.
If the stage-wise optimization:
\begin{equation}
\sup_{\substack{
S-T-X\\
I(S;T)\leq I(T;X)\\
S\in\mathsf M(X)
}}
I(S;X)
\end{equation}
is attained by $S^\star$, then $I(S^\star;T)=I(T;X)$.
\end{lemma}

\begin{IEEEproof}
We prove the lemma by forming a contradiction. Suppose, that $I(S^\star;T)<I(T;X)$.
Since $S^\star-T-X$,
\begin{equation}
I(T;X)
=
I(S^\star;X)+I(T;X|S^\star).
\end{equation}
If $I(T;X|S^\star)=0$, then $I(S^\star;X)=I(T;X)$,
which contradicts $I(S^\star;X)
\leq I(S^\star;T)
<I(T;X)$.
Hence $I(T;X|S^\star)>0$.

Because $T$ takes values in a standard Borel space, there exists a
finite-valued measurable function $Q=q(T)$ such that $I(Q;X|S^\star)>0$. 
Let $B\sim\operatorname{Bern}(\lambda)$ be independent of
$(S^\star,T,X)$, and define $\widehat S=(S^\star,W)$, where:
\begin{equation}
W=
\begin{cases}
Q,&B=1,\\
\mathtt e,&B=0.
\end{cases}
\end{equation}
Then $\widehat S-T-X$, and we have that:
\begin{align}
I(\widehat S;T)
&=
I(W,S^\star; T) \\
&= I(S^\star ; T) + I(W;T|S^\star) \\
&= I(S^\star ; T) + H(W|S^\star) - H(W|S^\star, T) \\ \notag
&= I(S^\star ; T) + \lambda H(Q|S^\star) + (1-\lambda)H(\mathtt e) \\
&\quad - [\lambda H(Q|S^\star, T) + (1-\lambda)H(\mathtt e)] 
\\
&= I(S^\star;T)
+
\lambda H(Q|S^\star),\\
I(\widehat S;X)
&=
I(S^\star;X)
+
\lambda I(Q;X|S^\star).
\end{align}
Since $Q$ is finite-valued,  $H(Q|S^\star)\leq\log_2|\mathcal Q|<\infty$. 

Let $\eta\triangleq I(T;X)-I(S^\star;T)>0$. We note that $H(Q|S^\star)>0$, since otherwise
$I(Q;X|S^\star)=0$. Choose $\lambda$ such that:
\begin{equation}
0<\lambda\leq
\min\left\{
1,\frac{\eta}{H(Q|S^\star)}
\right\}.
\end{equation}
Then
\begin{equation}
I(\widehat S;T)
=
I(S^\star;T)+\lambda H(Q|S^\star)
\leq I(T;X),
\end{equation}
whereas
\begin{equation}
I(\widehat S;X)
=
I(S^\star;X)+\lambda I(Q;X|S^\star)
>
I(S^\star;X).
\end{equation}

Now if $\widehat S\notin\mathsf M(X)$, replace it by a target-minimal
sufficient statistic $G=g(\widehat S)$ for $X$. Then $X-G-\widehat S$, and hence $I(G;X)=I(\widehat S;X)$ and $I(G;T)\leq I(\widehat S;T)$.
Thus $G$ is feasible and satisfies $I(G;X)=I(\widehat{S}; X) >I(S^\star;X)$. However, this contradicts the stage-wise optimality of $S^\star$. Therefore, $I(S^\star;T)=I(T;X)$.
\end{IEEEproof}

The following lemma demonstrates that if there are several terminal auxiliaries all achieving the optimal HCI value, then these optimal auxiliaries do not need to have much in common with each other. 

\begin{lemma}
\label{lem:independent_terminal_optimizers}
There exist finite-alphabet sources $X_1,X_2,X_3$ for which two
target-minimal terminal auxiliaries $T_2^\star$ and
$\widetilde T_2^\star$ both attain the HCI value, while $I(T_2^\star;\widetilde T_2^\star)=0$.
\end{lemma}

\begin{IEEEproof}
Let $A,B,Q$ be mutually independent
$\operatorname{Bern}(\tfrac12)$ random variables and define $X_1\triangleq(A,B)$, and $X_2=X_3\triangleq(Q,Y)$,
where:
\begin{equation}
Y\triangleq
\begin{cases}
A,&Q=0,\\
B,&Q=1.
\end{cases}
\end{equation}
We first choose $X_1$ as the reference.
Since $X_2=X_3$, the two orderings are equivalent. Moreover, $I(X_1;X_2)=1$.

Let $S$ be any feasible auxiliary at the $X_2$-stage. Then $S-(A,B)-(Q,Y)$ and $I(S;A,B)\leq 1$.
Since $Q$ is independent of $(S,A,B)$, we have that:
\begin{align}
I(S;X_2) 
&=I(S;Q,Y)\\
&=I(S;Q) + I(S;Y|Q) \\
&=I(S;Y|Q)\\
&=\frac12 I(S;A)+\frac12 I(S;B).
\end{align}
Because $A$ and $B$ are independent, we furthermore have that:
\begin{align}
I(S;A)+I(S;B)
&=
I(S;A,B) + I(A;B) -I(A;B|S)\\
&\leq I(S;A,B)\\
&\leq1.
\end{align}
Therefore, $I(S;X_2)\leq\frac12$.
Now consider $T_2^\star\triangleq A$ and $\widetilde T_2^\star\triangleq B$.
Both auxiliaries satisfy the MC constraint and saturate the IM
constraint:
\begin{equation}
I(A;X_1)=I(B;X_1)=1=I(X_1;X_2).
\end{equation}
Furthermore, $I(A;X_2)=I(B;X_2)=\frac12$. 
Since $X_3=X_2$, we have that:
\begin{align}
\min_{1\leq j\leq3}I(A;X_j)=\min_{1\leq j\leq3}I(B;X_j)=\frac12.
\end{align}
The missing reference cases are simpler and in a similar way it can be shown that using $X_2$ or $X_3$ as reference yields the fixed-reference value 1, which is greater than when using $X_1$ as the reference variable.  Hence, $R(X_1,X_2,X_3) = 1/2$ and both $A$ and $B$ attain it.
Nevertheless, because $A$ and $B$ are independent, $I(A;B)=0$. Therefore, even when the HCI is strictly positive, two terminal
optimizers do not necessarily share any mutual information.
\end{IEEEproof}

\begin{lemma}[Pairwise mutual-information upper bound]
\label{lem:pairwise_upper_bound}
Let $X_1,\ldots,X_n$ be arbitrarily distributed random variables.
Then
\begin{equation}
 R(X_1,\ldots,X_n)
\leq R_{\mathrm{ub}}(X_1,\ldots,X_n)\leq 
\min_{i\neq j} I(X_i;X_j).
\label{eq:pairwise_upper_bound}
\end{equation}
\end{lemma}

\begin{IEEEproof}
We have already establish that $R(X_1,\ldots,X_n) \leq R_{\mathrm{ub}}(X_1,\ldots,X_n)$. Thus, we only need to prove the last inequality. 
Fix a reference index $i$, an ordering $\pi^{(i)}$, and an arbitrary
feasible auxiliary chain:
\begin{equation}
T_1^{(i)}=X_i,\quad
T_2^{(i)},\ldots,T_{n-1}^{(i)}.
\end{equation}
The nested MC constraints imply, by composition, that the terminal
auxiliary satisfies:
\begin{equation}
T_{n-1}^{(i)}-X_i-\mathbf X.
\end{equation}
Thus, for every $j\neq i$, the data-processing inequality
gives:
\begin{equation}
I(T_{n-1}^{(i)};X_j)
\leq
I(X_i;X_j).
\end{equation}
Therefore, we must have that:
\begin{align}
\min_{1\leq k\leq n}
I(T_{n-1}^{(i)};X_k)
&\leq I(T_{n-1}^{(i)};X_j)\\
&\leq I(X_i;X_j).
\end{align}
Since this holds for every $j\neq i$, we finally get:
\begin{equation}
\min_k I(T_{n-1}^{(i)};X_k)
\leq
\min_{j\neq i}I(X_i;X_j).
\end{equation}
Taking the supremum over all feasible chains and then the maximum over
all orderings preserves the bound. 
\end{IEEEproof}

It is well-known that the mutual information is invariant under invertible transformations \cite{cover_thomas}. We can show that a similar result holds for the HCI. 
\begin{lemma}[Invariance under componentwise invertible transformations]
\label{lem:invertible_invariance}
Let $X_1,\ldots,X_n$ be arbitrarily distributed random variables.
For each $j$, let $Y_j=f_j(X_j)$, where $f_j$ is invertible, i.e., $g_j(Y_j)=X_j$ exists.
Then:
\begin{equation}
R(Y_1,\ldots,Y_n)
=
R(X_1,\ldots,X_n).
\label{eq:invertible_invariance}
\end{equation}
\end{lemma}

\begin{IEEEproof}
Let $\mathbf X=(X_1,\ldots,X_n)$ and $\mathbf Y=(Y_1,\ldots,Y_n)$. Then, for every auxiliary random variable $T$:
\begin{equation}
I(T;Y_j)=I(T;X_j),
\quad j=1,\ldots,n.
\label{eq:MI_invariance_component}
\end{equation}

Fix a reference index $i$, an ordering $\pi^{(i)}$, and a feasible
chain for $\mathbf X$:
\begin{equation}
T_1^{(i)}=X_i,\quad
T_2^{(i)},\ldots,T_{n-1}^{(i)}.
\end{equation}
Construct a corresponding chain for $\mathbf Y$ by setting:
\begin{equation}
\widehat T_1^{(i)}=Y_i,
\quad
\widehat T_\ell^{(i)}=T_\ell^{(i)},
\quad \ell=2,\ldots,n-1.
\end{equation}

We first verify the MC constraints. Since $X_i$ and $Y_i$ determine
one another almost surely, the first-stage Markov chain $T_2^{(i)}-X_i-\mathbf X$ is equivalent to $\widehat T_2^{(i)}-Y_i-\mathbf Y$.
For every subsequent stage, we similarly have that $T_{\ell+1}^{(i)}-T_\ell^{(i)}-\mathbf X$ 
is equivalent to $\widehat T_{\ell+1}^{(i)}
-\widehat T_\ell^{(i)}
-\mathbf Y$.
Thus, the transformed chain satisfies all MC constraints.
We also have that the mutual-information invariance gives:
\begin{align}
I(\widehat T_2^{(i)};\widehat T_1^{(i)})
&=I(T_2^{(i)};Y_i)\\
&=I(T_2^{(i)};X_i),
\end{align}
and
\begin{equation}
I\!\left(
\widehat T_1^{(i)};
Y_{\pi^{(i)}(1)}
\right)
=
I\!\left(
X_i;
X_{\pi^{(i)}(1)}
\right).
\end{equation}
Hence the first IM constraint for the $\mathbf Y$-chain follows from
the corresponding IM constraint for the $\mathbf X$-chain.

For every $\ell\geq2$,
\begin{equation}
I(\widehat T_{\ell+1}^{(i)};\widehat T_\ell^{(i)})
=
I(T_{\ell+1}^{(i)};T_\ell^{(i)}),
\end{equation}
while
\begin{equation}
I\!\left(
\widehat T_\ell^{(i)};
Y_{\pi^{(i)}(\ell)}
\right)
=
I\!\left(
T_\ell^{(i)};
X_{\pi^{(i)}(\ell)}
\right).
\end{equation}
Therefore, all subsequent IM constraints are also preserved. Moreover, all mutual information expressions are preserved.
Thus, every feasible chain for $\mathbf X$ induces a feasible chain
for $\mathbf Y$ with the same terminal value and vice versa. It follows that:
\begin{equation}
R(Y_1,\ldots,Y_n)
=
R(X_1,\ldots,X_n).
\end{equation}
\end{IEEEproof}

\begin{lemma}[Vanishing under pairwise independence]
\label{lem:R_zero_one_pair}
Let $X_1,\ldots,X_n$ be arbitrarily distributed random variables. If
there exist indices $i\ne j$ such that $I(X_i;X_j)=0$,
then $R(X_1,\ldots,X_n)=0$.
\end{lemma}

\begin{IEEEproof}
By the pairwise upper bound, $R(X_1,\ldots,X_n)
\le
\min_{k\ne \ell}I(X_k;X_\ell)
\le I(X_i;X_j)=0.$
Moreover, since mutual information is nonnegative, the HCI is also nonnegative. 
\end{IEEEproof}

\begin{lemma}[Entropy upper bound]
\label{lem:upperbound}
Let $n\ge2$, and let $X_1,\ldots,X_n$ be discrete random variables.
Then
\begin{equation}
0\leq R(X_1,\ldots,X_n)
\le
\min_{1\le i\le n}H(X_i).
\end{equation}
\end{lemma}

\begin{IEEEproof}
The upper bound follows from the pairwise upper bound in Lemma~\ref{lem:pairwise_upper_bound}, and the that fact that $I(X_i ;X_j) \leq \min ( H(X_i), H(X_j))$ for discrete $X_i,X_j$. The lower bound follows since the mutual information is non-negative.
\end{IEEEproof}

\begin{lemma}\label{lem:HX}
Let $X_1,\ldots,X_n$ be discrete random variables satisfying $X_1=\cdots=X_n$ and $H(X_1)<\infty$. Then
\begin{equation}
R(X_1,\ldots,X_n)=H(X_1).
\end{equation}
\end{lemma}

\begin{IEEEproof}
The identity auxiliary satisfies all the constraints in \eqref{eq:stage_feasible_set}. Moreover, it is stage-wise optimal. 
Indeed, every feasible
auxiliary $S$ satisfies for any $\ell$:
\begin{equation}
I(S;X_{\pi^{(i)}(\ell)})=I(S;X)\leq H(X),
\end{equation}
whereas the choice $S=X$ attains $H(X)$.
It follows that for any reference variable, let $T_1^{(i)}=X_i=X$ and at any stage $\ell>2$ let $T_\ell^{(i)} = T_{\ell-1}^{(i)} = X_i$. 
Its terminal value is:
\begin{equation}
\min_{1\leq j\leq n}
I(T_{n-1}^{(i)};X_j)
=
\min_{1\leq j\leq n}I(X;X_j)
=H(X).
\end{equation}
Since this holds for every reference variable and every ordering, $R(X_1,\ldots,X_n)= H(X)$.
\end{IEEEproof}

\begin{lemma}[Independent-component representation]
\label{lem:independent_components}
Let $Z_1,\ldots,Z_k$ be mutually independent discrete random variables
satisfying $0\leq H(Z_j)<\infty, j=1,\ldots,k$. For each $i\in\{1,\ldots,n\}$, let
$A_i\subseteq\{1,\ldots,k\}$ and define $X_i\triangleq (Z_j)_{j\in A_i}$.
Then
\begin{equation}
R(X_1,\ldots,X_n)
=
H\!\left((Z_j)_{j\in\bigcap_{i=1}^nA_i}\right)
=
\sum_{j\in\bigcap_{i=1}^nA_i}H(Z_j).
\end{equation}
\end{lemma}

\begin{IEEEproof}
Fix a stage and suppose that the current representation $T$ is
equivalent to $Z_B\triangleq(Z_j)_{j\in B}$.
Let the current target be $X_a=Z_{A_a}$,
and define $C\triangleq Z_{B\cap A_a}$.
By independence of the components, $I(T;X_a)=H(C)$.

For every feasible $S$, the MC and IM constraints give (since $S$ is a (stochastic) mapping of $T$):
\begin{equation}
I(S;X_a)
\leq I(S;T)
\leq I(T;X_a)
=
H(C).
\end{equation}
The choice $S=C$ is feasible and attains the upper bound $I(C;T)=I(C;X_a)=H(C)$.
Moreover, $C\in\mathsf M(X_a)$, because $C$ is a deterministic
function of $X_a$. Hence, the maximum stage-wise relevance is $H(C)$.
Let $S^\star$ be any target-minimal stage-wise optimizer. Then, it must attain the upper bound:
\begin{equation}
I(S^\star;X_a)
=
I(S^\star;T)
=
H(C).
\end{equation}
Since $S^\star-T-X_a$, it follows that:
\begin{equation}
I(S^\star;T)
=
I(S^\star;X_a)+I(S^\star;T|X_a).
\end{equation}
The above further implies that $I(S^\star;T|X_a)=0$, which means that there cannot be be private randomness in $S^*$ about $T$.
Writing $T=(C,D)$ and $X_a=(C,E)$, where $C,D,E$ consist of disjoint collections of the independent
components, gives $I(S^\star;T|C)=0$. It follows that $I(S^\star;C) = I(S^\star;T) = H(C)$,
and hence $H(C|S^\star)=0$. Thus, $C$ is a deterministic function of $S^\star$. Furthermore,
$I(S^\star;T|C)=0$ and $S^\star-T-X_a$ imply $X_a-C-S^\star$.
Since $S^\star\in\mathsf M(X_a)$, target minimality implies that
$S^\star$ is also a deterministic function of $C$. Thus, at every stage, the optimizer $S^\star$ is
equivalent to the components $C$ shared by the current representation and
the current target. This holds for any reference $X_i=Z_{A_i}$ and ordering $\pi^{(i)}$, which means that:
\begin{equation}
\min_j I(T_{n-1}^{(i)};X_j)
=
H\!\left(
Z_{\bigcap_{r=1}^nA_r}
\right).
\end{equation}
The right hand side is independent of the reference and ordering, which means that:
\begin{align}
R(X_1,\ldots,X_n)
&=
H\!\left(
(Z_j)_{j\in\bigcap_{i=1}^nA_i}
\right) = \sum_{j\in\bigcap_{i=1}^nA_i}H(Z_j).
\end{align}
\end{IEEEproof}

We will now establish a monotonicity result for the HCI. It turns out to be simple to prove that the upper bound $R_{\mathrm{ub}}$ in \eqref{eq:Rub} satisfies $R_{\mathrm{ub}}(X_1,\dotsc, X_{n-1}) \geq R_{\mathrm{ub}}(X_1,\dotsc, X_{n})$. While such a result must intuitively hold also for the HCI, we have not been able to prove it without including additional restrictions on the feasible set of auxiliaries. In particular, if we further impose that the optimal auxiliaries are so-called universal source dominating, then we can establish the desired monotonicity result as we show by the next lemma. 

\begin{lemma}[Monotonicity under universal source dominance]
\label{lem:HCI_monotonicity}
Let $X_1,\ldots,X_n$ be arbitrarily distributed random variables.
Assume that, at every stage of the HCI construction, there exists a
target-minimal stage-wise optimizer that is universally
source-dominating. Specifically, for a current representation $T$,
current target $X_a$, and budget $\Delta=I(T;X_a)$,
assume that there exists a stage-wise optimizer $S^\star$ such that,
for every feasible auxiliary $U$ satisfying $U-T-\mathbf X$ and $I(U;T)\leq\Delta$, $U\in\mathsf M(X_a)$,
there exists a stochastic degradation $\widetilde U$ of $S^\star$
for which:
\begin{equation}
\widetilde U-S^\star-\mathbf X
\end{equation}
and
\begin{equation}
(\widetilde U,\mathbf X)
\stackrel{d}{=}
(U,\mathbf X).
\end{equation}
Then
\begin{equation}
R(X_1,\ldots,X_n)
\leq
R(X_1,\ldots,X_{n-1}).
\label{eq:HCI_monotonicity}
\end{equation}
\end{lemma}

Before proving the lemma, we emphasize that universal source dominance
does not require the original auxiliary $U$ to satisfy $U-S^\star-\mathbf X$
under its given joint coupling with $S^\star$ and $\mathbf X$.
Instead, it only requires the existence of an auxiliary
$\widetilde U$, generated by stochastically degrading $S^\star$,
such that $\widetilde U-S^\star-\mathbf X$ and $(\widetilde U,\mathbf X)\stackrel d=(U,\mathbf X)$ \cite{Blackwell1953}.

\begin{IEEEproof}
Let $\mathbf X^{(n)}=(X_1,\ldots,X_n)$ and $\mathbf X^{(n-1)}=(X_1,\ldots,X_{n-1})$.
Fix a reference index $i\in\{1,\ldots,n-1\}$, an ordering
$\pi'$ of the remaining sources in the $n$-source construction $\mathbf X^{(n)}$, and
a stage-wise-optimal auxiliary chain corresponding to this reference
and ordering. Let $\pi'_{-n}$ denote the ordering obtained from
$\pi'$ by deleting $X_n$. We compare this chain with the $(n-1)$-source construction having
reference $X_i$ and ordering $\pi'_{-n}$. From the first source and until the position at
which $X_n$ occurs in $\pi'$, the two constructions process the same
sources in the same order and solve the same stage-wise optimization
problems. We may therefore choose identical stage-wise optimizers until the point where $X_n$ is encountered.
Let $X_n$ be the source at position $\ell$ in the ordering $\pi'$, i.e., $\pi'(\ell)=n$, then the representation
immediately before the $X_n$-stage is $T=T_\ell^{(i)}$. We assume $\ell<n-2$ otherwise $X_n$ is processed last and the result follows immediately,  since the sequence of auxiliaries obtained for the two source chains are equivalent up till the last element. Including additional auxiliaries in the end can only decrease the resulting mutual information. We therefore proceed by assuming that 
 $X_n$ is processed at a nonterminal position and
is followed by a source $X_a$ from the reduced collection, where
$a\in\{1,\ldots,n-1\}$. Let $V$ denote the representation obtained
by processing $X_n$ from $T$, and let $U$ denote the subsequent
representation obtained by processing $X_a$ from $V$. The relevant subset of the whole $n$-source chain is therefore:
\begin{equation}
T\longrightarrow V\longrightarrow U,
\end{equation}
where the first transition corresponds to $X_n$ and the second to
$X_a$. On the other hand, in the $(n-1)$-source construction, $X_n$ is absent, so
$X_a$ is processed directly from $T$ without going through $V$.

By the data processing inequality and the IM constraint at the $X_a$-stage, we have that:
\begin{align}
I(U;T)
&\leq I(U;V)\\
&\leq I(V;X_a)\\
&\leq I(T;X_a).
\label{eq:HCI_bypass_rate}
\end{align}
The last inequality follows from $V-T-X_a$.

The two stochastic kernels used in the longer chain, $P_{V|T}$ and $P_{U|V}$,
can be composed by marginalizing over $V$ to form the direct kernel:
\begin{equation}
P_{\bar U|T}(u|t)
\triangleq
\int P_{U|V}(u|v)\,P_{V|T}(dv|t).
\label{eq:composed_kernel}
\end{equation}
Because the longer chain satisfies $U-V-T-\mathbf X$, its joint law
factors as:
\begin{equation}
P_{\mathbf X,T,V,U}
=
P_{\mathbf X,T}P_{V|T}P_{U|V}.
\end{equation}
Marginalizing over $V$ yields:
\begin{equation}
P_{\mathbf X,T,U}
=
P_{\mathbf X,T}
\int P_{U|V}\,dP_{V|T},
\end{equation}
where the integral is identical to the composed kernel $P_{\bar U|T}$ of \eqref{eq:composed_kernel}.
It follows that:
\begin{equation}
(\bar U,T,\mathbf X)\stackrel d=(U,T,\mathbf X).
\end{equation}
Thus, eliminating the intermediate representation $V$ changes only
the factorization of the channel, not the joint source experiment
induced by the terminal auxiliary.
The auxiliary $\bar U$ generated from $T$ using this kernel has the
same joint distribution with the source vector as $U$:
\begin{equation}
(\bar U,\mathbf X^{(n-1)})
\stackrel d=
(U,\mathbf X^{(n-1)}).
\end{equation}
Moreover, $\bar U-T-\mathbf X^{(n-1)}$. We therefore have that $I(\bar U;T) = I(U;T) \leq I(T;X_a)$,
so $\bar U$ satisfies the IM constraint for the direct
$T$-to-$X_a$ stage. Finally, because $(\bar U,X_a)\stackrel d=(U,X_a)$ and $U\in\mathsf M(X_a)$, we also have $\bar U\in\mathsf M(X_a)$.
Thus, $\bar U$ is a feasible candidate when $X_a$ is processed
directly from $T$ in the shorter $(n-1)$-source construction.

Let $S^\star$ be a universally source-dominating stage-wise optimizer
for this direct $X_a$-stage. By universal dominance, there exists a
stochastic degradation $\widetilde U$ of $S^\star$ satisfying $(\widetilde U,\mathbf X^{(n-1)}) \stackrel d= (U,\mathbf X^{(n-1)}).$
Hence the shorter representation $S^\star$ can reproduce the complete
source experiment induced by the longer representation $U$. This holds for any stage. Thus, let $T_{\mathrm{long}}$ and $T_{\mathrm{short}}$ denote the resulting terminal auxiliaries. Then the longer terminal source experiment is a stochastic degradation of the shorter one, so:
\begin{equation}
I(T_{\mathrm{long}};X_j)
\leq
I(T_{\mathrm{short}};X_j),
\quad j=1,\ldots,n-1.
\label{eq:HCI_terminal_dominance}
\end{equation}
It follows that:
\begin{align}
\min_{1\leq j\leq n}I(T_{\mathrm{long}};X_j)
&\leq
\min_{1\leq j\leq n-1}I(T_{\mathrm{long}};X_j)\\
&\leq
\min_{1\leq j\leq n-1}I(T_{\mathrm{short}};X_j).
\label{eq:HCI_terminal_bound}
\end{align}
This holds for every reference $i$ and $n$-source ordering $\pi'$. It therefore also holds when maximizing over the orderings, which establishes that:
\begin{align}
R(X_1,\ldots,X_n)
&\leq R(X_1,\ldots,X_{n-1}),
\end{align}
and thereby proves \eqref{eq:HCI_monotonicity}.
\end{IEEEproof}

\section{HCI for Gaussian Sources}
\label{sec:gaussian}
In this section we focus on jointly Gaussian sources and find $R(X_1,\dotsc, X_n)$ for any $n\geq 2$. 
Before presenting the main theorem, we introduce a notion of order statistics of correlations, which will be needed in the sequel.  
Let $X_1,\dotsc, X_n$ be jointly distributed random variables. 
Let $\rho_{j,k}\triangleq \mathrm{corr}(X_j,X_k)  =\frac{\mathrm{Cov}(X_j,X_k)}
{\sqrt{\mathrm{Var}(X_j)\mathrm{Var}(X_k)}}, j\neq k$, be the correlation coefficient between $X_j$ and $X_k$. 
Fix $i\in\{1,\dots,n\}$ and consider the set
$
\{|\rho_{i,j}| : j\neq i\},
$ which contains $n-1$ elements.
The \emph{order statistics} $
r_{i,(1)} \le r_{i,(2)} \le \cdots \le r_{i,(n-1)}$
are defined as the elements of the set $\{|\rho_{i,j}| : j\neq i\}$
arranged in non-decreasing order with 
with $r_{i,(1)}=\min_{j\neq i}|\rho_{i,j}|$ and
$r_{i,(n-1)}=\max_{j\neq i}|\rho_{i,j}|$.

\begin{theorem}[HCI for Gaussian sources]
\label{thm:general_gaussian}
Let $X_1,\dots,X_n$ be zero-mean scalar jointly Gaussian random variables with correlation coefficients $\rho_{j,k}\triangleq \mathrm{corr}(X_j,X_k), j\neq k$.
For fixed $i$, let
$r_{i,(1)}\le r_{i,(2)}\le\cdots\le r_{i,(n-1)}$
be the order statistics of $\{|\rho_{i,j}|:j\neq i\}$.
Then, the HCI of $X_1,\dots,X_n$ is given by:
\begin{equation}
R(X_1,\dots,X_n)
=
\min_{1\le i\le n}
-\frac12\log_2\!\left(
1-
r_{i,(1)}^2
\prod_{m=2}^{n-1}
r_{i,(m)}^{\,2^{\,m-1}}
\right),    
\end{equation}
where the empty product is interpreted as one.
\end{theorem}
\begin{IEEEproof}
    See the appendix.
\end{IEEEproof}

\begin{example}
\label{ex:n3_nonsymmetric}
Let $(X_1,X_2,X_3)$ be jointly Gaussian with zero mean and covariance matrix:
\begin{equation}
\Sigma =
\begin{pmatrix}
4 & 1 & 0.5 \\
1 & 1 & 0.3 \\
0.5 & 0.3 & 2
\end{pmatrix}.
\end{equation}
The pairwise correlations are $
\rho_{12}=0.5,
\rho_{13}\approx 0.177,
\rho_{23}\approx 0.212.
$
The pairwise mutual informations are $I(X_1;X_2)\approx 0.2075,
I(X_1;X_3)\approx 0.0229,
I(X_2;X_3)\approx 0.0332$.
For $n=3$, Theorem~\ref{thm:general_gaussian} reduces to: 
\begin{equation}
R(X_1,X_2,X_3) 
= \min_{i=1,2,3} 
-\frac12\log_2\!\Bigl(1-r_{i,(1)}^2\,r_{i,(2)}^2\Bigr),    
\end{equation}
where $r_{i,(1)}\le r_{i,(2)}$ are the order statistics of
$\{|\rho_{i,j}|:j\neq i\}$. 
For reference $i=1,2,$ and $3$, we have the order
statistics $
(r_{1,(1)}=0.177,\; r_{1,(2)}=0.5),
(r_{2,(1)}=0.212,\; r_{2,(2)}=0.5),
(r_{3,(1)}=0.177,\; r_{3,(2)}=0.212),
$
respectively.
This leads to 
\begin{equation}
R(X_1,X_2,X_3) =
-\tfrac12\log_2\!\bigl(1-(0.177)^2(0.212)^2\bigr)
\approx 0.0010.
\end{equation}
Although $X_3$ does not have the largest nor the smallest variance, it is optimal as the reference
variable because it yields the smallest product of absolute correlations, i.e., it is the variable whose product of correlations to the others is the weakest.  
\end{example}
\hfill$\triangle$

\begin{corollary}[Equicorrelated Gaussian variables]
\label{cor:equal_gauss}
Let $X_1,\dots,X_n$ be zero-mean jointly equicorrelated Gaussian random variables with correlation coefficient $\rho$. 
Then, we obtain from Theorem~\ref{thm:general_gaussian} that:
\begin{equation}
    R(X_1,\dotsc, X_n) = -\frac12\log_2\!\Bigl(1-\rho^{\,2^{\,n-1}}\Bigr).
\end{equation}
\end{corollary}
\hfill$\triangle$

\begin{theorem}
\label{thm:Gaussian_n3}
Let $X_1,X_2,X_3$ be jointly Gaussian scalar random variables as in
Theorem~\ref{thm:general_gaussian}. Then,
\begin{equation}
R_{\mathrm{ub}}(X_1,X_2,X_3)
=
R(X_1,X_2,X_3).
\end{equation}
\end{theorem}
\begin{IEEEproof}
    See the appendix.
\end{IEEEproof}

Fig.~\ref{fig:gaussian_cmi} shows the HCI from Corollary~\ref{cor:equal_gauss} as a function of the correlation coefficient $\rho$ for $n=2,3,$ and $4$. Note that we must have $-1/(n-1)<\rho \leq 1$. It can be seen that the HCI is symmetric in $\rho$ (around $\rho=0$) in its valid range. For comparison, we have also shown the interaction information $II(X_1;X_2|X_3)$, which is defined as \cite{Fano:1961,McGill}:
\begin{align}
    II(X_1;X_2|X_3) &= I(X_1;X_2) - I(X_1;X_2|X_3) \\
    &= -\frac{1}{2}\log_2\!\left(
\frac{(1-\rho)(1+\rho)^3}{1+2\rho}
\right).
\end{align}
It can be seen that $II(X_1;X_2|X_3)$ becomes negative for $\rho<0$ and is therefore not symmetric around $\rho=0$. Moreover, for $\rho>0$, the following sandwich structure can be observed: $R(X_1,X_2) \geq II(X_1;X_2|X_3) \geq R(X_1,X_2,X_3)$. It is easy to prove that this holds for any $1>\rho\geq 0$. Clearly, $I(X_1;X_2) - I(X_1;X_2|X_3) \leq I(X_1;X_2)$. Thus, the left hand side of the sandwich trivially holds. 
To prove the right hand side, we can write out the explicit expressions, which means that we should show that:
\begin{align}
  -\frac12\log_2\!\left(
\frac{(1-\rho)(1+\rho)^3}{1+2\rho}
\right) \geq
-\frac12\log_2(1-\rho^4).
\end{align}
Since $-\frac12\log_2(\cdot)$ is decreasing, this is equivalent to:
\begin{equation}
\frac{(1-\rho)(1+\rho)^3}{1+2\rho}
\leq
1-\rho^4.
\end{equation}
Using that $1-\rho^4
=
(1-\rho)(1+\rho)(1+\rho^2)$,
we obtain:
\begin{align}
\frac{(1-\rho)(1+\rho)^3}{1+2\rho}
&\leq
(1-\rho)(1+\rho)(1+\rho^2).
\end{align}
For $0\leq \rho <1$, we may divide by $(1-\rho)(1+\rho)>0$, yielding:
\begin{equation}
(1+\rho)^2
\leq
(1+2\rho)(1+\rho^2).
\end{equation}
Expanding both sides gives:
\begin{equation}
1+2\rho+\rho^2
\leq
1+2\rho+\rho^2+2\rho^3,
\end{equation}
which is equivalent to:
\begin{equation}
0\leq 2\rho^3.
\end{equation}
This holds whenever $\rho\geq 0$, which proves the sandwich.

\begin{figure}[t!]
        \centering
 \includegraphics[width=9cm]{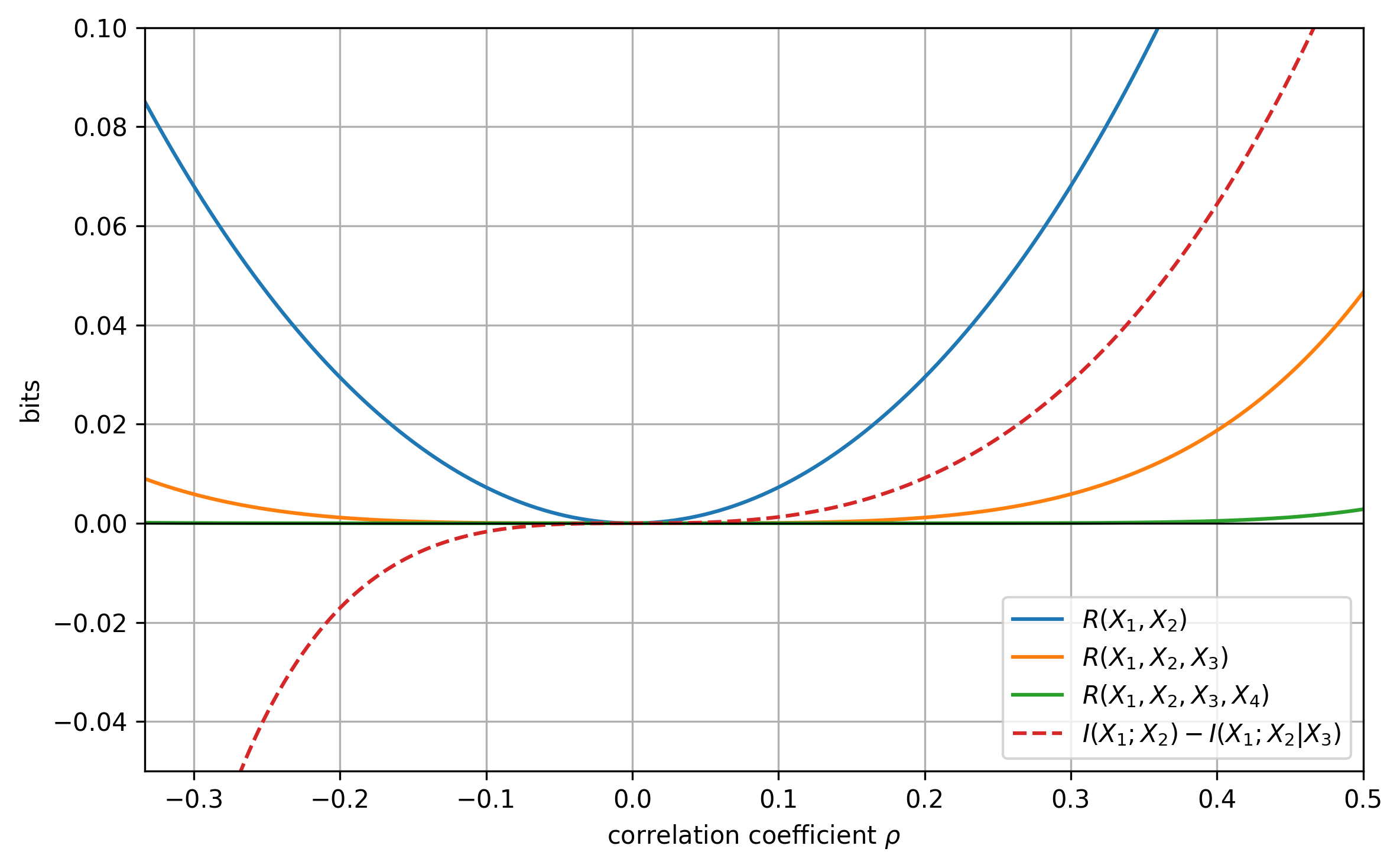}        
        \caption{The HCI for equicorrelated Gaussian random variables. For comparison, we have also shown the interaction information.}
        \label{fig:gaussian_cmi}
  \end{figure}

\section{Common Information for Discrete Sources}
\label{sec:discrete}

In this section, we present results for some discrete sources, where we are able to find closed-form expressions for $R(X_1,\dotsc, X_n)$.
We begin by the following corollary whose proof follows directly from Lemma~\ref{lem:independent_components}.
\begin{corollary}
\label{theo:Rnprime_allL}
Let $n>2$ and let $\mathcal Z=\{Z_1,\dots,Z_n\}$ be mutually independent and identically distributed
discrete random variables with $0<H(Z)<\infty$.
For each $k\in \{1,\dotsc, n\}$ define:
\begin{equation}
X_k \triangleq (Z_1,\dots,Z_{k-1},Z_{k+1},\dots,Z_n),    
\end{equation}
i.e., $X_k$ contains all $Z$-components except $Z_k$.
Fix any index set $\mathcal I\subseteq \{1,\dotsc, n\}$ with $|\mathcal I|=n'$ and assume $1\le n'\le n$.
Let $\mathbf X^{(n')}\triangleq \{X_i:i\in\mathcal I\}$ denote the selected $n'$ variables.
Then, the HCI among $\mathbf X^{(n')}$ is:
\begin{equation}
\label{eq:Rnprime_allL}
R(\mathbf X^{(n')}) = (n-n')H(Z).
\end{equation}

\hfill$\triangle$
\end{corollary}

For the collection $\mathbf X^{(n')}$ considered here, the maximal common
deterministic component is given by
$W=(Z_k)_{k\in[n]\setminus\mathcal I}$, and it follows from the characterization
in~\cite{gacs} that:
\begin{equation}
C_{\mathrm{GK}}(\mathbf X^{(n')}) = H(W) = (n-n')H(Z) = R(\mathbf X^{(n')}).    
\end{equation}

It is not always the case that the HCI coincides with the G\'acs--K\"orner common information for discrete sources. For example, below we consider correlated Bernoulli sources that do not share any deterministic structure, which implies that the G\'acs--K\"orner common information is zero \cite{gacs}. However, since they are statistically dependent upon each other,  their HCI is non-zero. 

\begin{theorem}[HCI for binary common-source variables]
\label{thm:Bernoulli_HCI}
Let $U\sim\mathrm{Bern}(\tfrac12)$ and let
$N_1,\ldots,N_n$ be mutually independent with
$N_j\sim\mathrm{Bern}(p_j)$, where $0\le p_j\le \tfrac12$.
Define $X_j = U\oplus N_j, j=1,\ldots,n,$
and let $c_j \triangleq 1-2p_j \in[0,1]$.
For each reference index $i$, let $d_{i,(1)}\le d_{i,(2)}\le \cdots \le d_{i,(n-1)}$
be the order statistics of the set $\{c_j:j\neq i\}$. Then
\begin{equation}
R(X_1,\ldots,X_n)
=
\min_{1\le i\le n}
\left[
1-h_2\!\left(\frac{1-\beta_i}{2}\right)
\right],    
\end{equation}
where
\begin{equation}
\beta_i
=
d_{i,(1)}\,
c_i^{2^{n-2}}
\prod_{m=2}^{n-1}
d_{i,(m)}^{\,2^{m-2}} .
\end{equation}

\hfill$\triangle$
\end{theorem}

\begin{remark}
    For $n=2$, the empty product is interpreted as one, so $\beta_i=c_ic_j, j\neq i$,
and hence:
\begin{equation}
R(X_1,X_2)=1-h_2\!\left(\frac{1-c_1c_2}{2}\right)
=I(X_1;X_2).    
\end{equation}
\end{remark}

\begin{corollary}[HCI for equicorrelated binary common-source variables]
\label{cor:equicorrelated_Bernoulli}
Let $X_1,\dotsc, X_n$ be given as in Theorem~\ref{thm:Bernoulli_HCI} but where $p_j=p, \forall j,$ so that $N_1,\dotsc, N_n$  are 
 mutually independent and identically distributed according to $N_j\sim\mathrm{Bern}(p), 0\leq p\leq\tfrac12.$
Let $c\triangleq1-2p$.
Then
\begin{equation}
R(X_1,\ldots,X_n)
=
1-h_2\!\left(
\frac{1-c^{\,2^{n-1}}}{2}
\right).
\label{eq:equicorrelated_Bernoulli_HCI}
\end{equation}
\hfill$\triangle$
\end{corollary}

\begin{figure}[t!]
    \centering
        \centering
 \includegraphics[width=9cm]{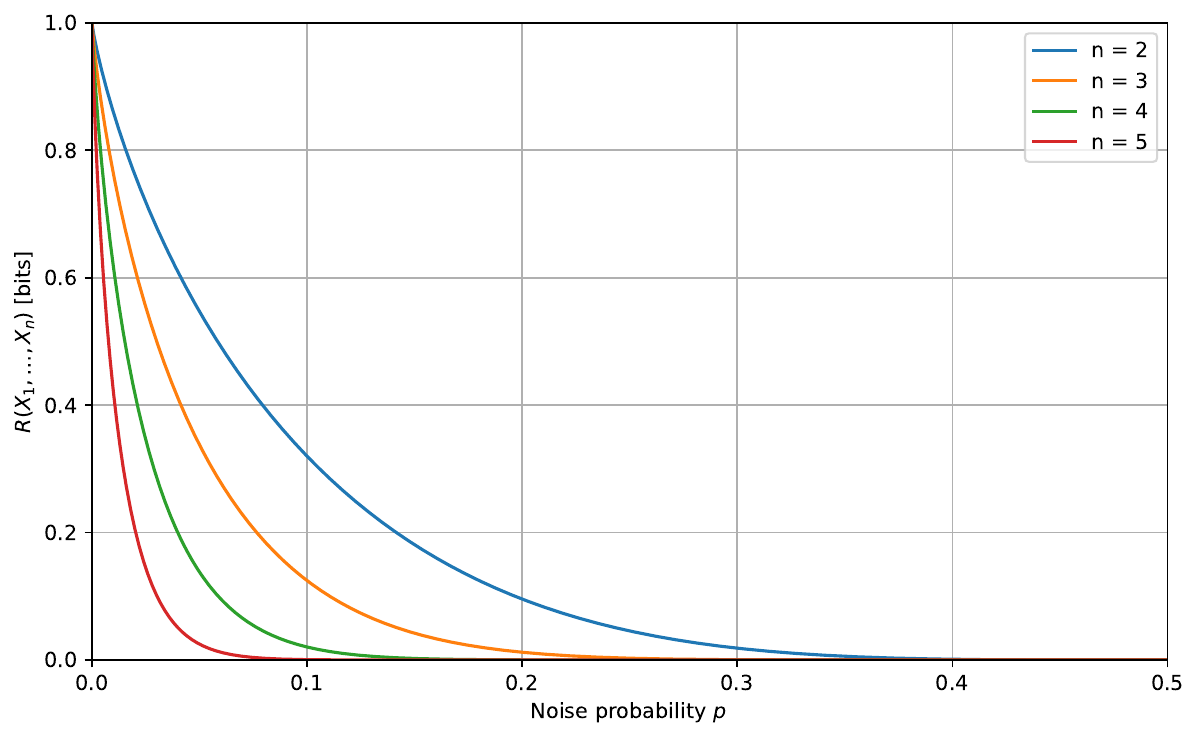}        
        \caption{Higher-order common information $R(X_1,\dotsc, X_n)$ for $n=2,3,4,5$, equicorrelated Bernoulli sources as given by Theorem~\ref{thm:Bernoulli_HCI}.}
        \label{fig:bernoulli_cmi}
  \end{figure}

The HCI $R(X_1,\dotsc, X_n)$ for the equicorrelated Bernoulli model is illustrated
in Fig.~\ref{fig:bernoulli_cmi} as a function of the noise parameter
$0\le p\le \tfrac12$, where $N_i\sim\mathrm{Bern}(p)$.
 When $p=0$, there is no noise and all variables coincide
($X_1=\cdots=X_n = U$), implying complete overlap of information and
$R(X_1,\dotsc, X_n)=R(X_1,X_2)=H(U)=1$ for all $n$. At the opposite extreme, when $p=\tfrac12$,
the variables are independent and the common information vanishes, yielding
$R(X_1,\dotsc, X_n)=R(X_1,X_2)=0$. For intermediate values $0<p<\tfrac12$, the common information is
strictly positive and decreases monotonically with $n$:
\begin{align}
    0&<R(X_1,\dotsc, X_n)<R(X_1,\dotsc, X_{n-1}) \\
    &<\cdots<R(X_1,X_2)<H(U) = 1.
\end{align}

\section{Approximating Common Information for General Sources}
\label{sec:general_sources}

In this section, we consider general sources and propose a practical
procedure for constructing an HCI-inspired approximation from
finite-length data. The procedure is motivated by the iterative
information-bottleneck construction, in which each auxiliary must
satisfy the Markov-chain, information-matching, and target-minimality
constraints while maximizing relevance to the current target source.
Our approximation uses the additive Gaussian test channels that are
optimal for jointly Gaussian sources. At each stage, the auxiliary is
constructed by adding independent zero-mean Gaussian noise to the
current representation. The noise variance is selected so that the
information-matching constraint in \eqref{eq:stage_feasible_set} is approximately saturated:
\begin{equation}
I(T_{\ell+1};T_\ell)
\approx
I(T_\ell;X_{\pi(\ell)}).
\end{equation}
This choice is also motivated by
Lemma~\ref{lem:IM_saturation}, which shows that, when the stage-wise
maximum is attained and the IM budget is finite, every stage-wise
maximizer must saturate the IM constraint.
Saturation is, however, only a necessary condition for stage-wise
optimality. For non-Gaussian sources, an additive Gaussian channel need
not maximize the source-relevance $I(T_{\ell+1};X_{\pi(\ell)})$ among all feasible stochastic kernels, nor is target minimality
automatically guaranteed. In addition, mutual-information estimation
from finite data permits only approximate enforcement of the IM
constraint. Thus, the resulting quantity should be interpreted as an HCI-inspired numerical approximation
that is exact for the Gaussian models covered by
Theorem~\ref{thm:general_gaussian}, subject to accurate estimation and
optimization of the noise variance.

We will provide an example with real-world EEG data and demonstrate that for $n=3$, the HCI-inspired statistic is significantly associated with the neural response variables beyond which can be explained by the minimum of the pairwise mutual information.
As a real-world application, we consider the 64-channel scalp EEG dataset described in \cite{fuglsang:2017}. In this auditory attention experiment, normal-hearing subjects were simultaneously presented with two competing speech streams originating from spatially separated acoustic sources. Subjects were instructed to attend to one speech stream (the target) while ignoring the other (the distractor).
For each trial, the temporal envelopes \cite{Lorenzi:2007} of the target and distractor speech signals, denoted by ($\mathrm{env}_T$) and ($\mathrm{env}_D$), respectively, were extracted and downsampled to 64 Hz to match the EEG sampling frequency. The EEG preprocessing included removal of ocular artifacts and exclusion of channels with poor signal quality. After preprocessing, data from 15 subjects were retained, each contributing 60 trials. All EEG signals and acoustic envelopes were normalized to have zero mean and unit variance.
Following the extended international 10--20 electrode placement system \cite{Niedermeyer:2004}, we focus on the temporal electrodes ($\mathrm{FT}_7,\mathrm{T}_7$), which are located near the left auditory cortex. 

\subsection{Computing Pairwise and Higher-Order Common Information}

For each trial of each subject, we computed the empirical mutual information using the corresponding EEG and acoustic envelope time series. Specifically, we estimated $I(\mathrm{FT}_7;\mathrm{T}_7)$, $I(\mathrm{FT}_7;\mathrm{env}_T)$,
$I(\mathrm{FT}_7;\mathrm{env}_D)$,
$I(\mathrm{T}_7;\mathrm{env}_T)$, and 
$I(\mathrm{T}_7;\mathrm{env}_D)$,
using the IDTxl mutual-information estimator \cite{Wollstadt2019IDTxl}.\footnote{We set the number of nearest neighbors in the knn-based estimator to 6.}
We define $R_2^T$ and $R_2^D$ as the minimum pair-wise mutual information related to the target and distractor stimuli, respectively:
\begin{align}
R_2^T
&=\min\{I(\mathrm{FT}_7;\mathrm{env}_T), I(\mathrm{T}_7;\mathrm{env}_T),I(\mathrm{FT}_7;\mathrm{T}_7)\}, \\
R_2^D
&=\min\{I(\mathrm{FT}_7;\mathrm{env}_D), I(\mathrm{T}_7;\mathrm{env}_D),I(\mathrm{FT}_7;\mathrm{T}_7)\}.
\end{align}
Similarly, we define $R_3^T$ and $R_3^D$ as approximations to the  third-order common information measures 
$R_3^T \approx R(\mathrm{FT}_7,\mathrm{T}_7,\mathrm{env}_T)$, and $R_3^D \approx R(\mathrm{FT}_7,\mathrm{T}_7,\mathrm{env}_D)$,
where $R_3^T$ and $R_3^D$ were estimated as follows. To compute $R_3^T$ we first fixed a reference variable, say $X_1=\mathrm{FT}_7$. Then, we fixed  a permutation $\pi$ of the remaining variables, e.g., 
$X_2=\mathrm{T}_7, X_3=\mathrm{env}_T$. The auxiliary variables are obtained as $T_1 = X_1$, and $T_2=T_1 + N$ by adding i.i.d., zero-mean Gaussian distributed noise with variance $\sigma^2$ to each element in the time series $T_1$. The variance $\sigma^2$ was adjusted so that $|I(T_2;X_1) - I(X_1;X_2)| < 0.1 I(X_1;X_2).$  
Using $T_2$ as a degraded representation of $X_1$ we compute $I(T_2;X_3)$. To reduce the Monte Carlo sampling variance, we repeat this procedure 10 times using independent realizations of $N$, and the median value of $I(T_2;X_3)$ over these realizations is retained. For each fixed reference variable, the procedure is repeated for the orderings of the remaining variables. Finally, the approximation to the third-order common information is obtained by minimizing over the reference variable and maximizing over the orderings for each reference variable. 

To summarize and formalize the procedure, let $\mathbf X=(X_1,X_2,X_3)$ denote the variables in the triplet under consideration. For each reference variable $X_i$, let $S_2^{(i)}$ denote the set of all permutations of $\mathcal I_i=\{1,2,3\}\setminus\{i\}$.
For a given permutation $\pi^{(i)}\in S_2^{(i)}$, we construct the degraded
auxiliary variable:
\begin{equation}
T_2^{(i)} = X_i + N,    
\end{equation}
where $N$ is an i.i.d.\ zero-mean Gaussian noise process whose variance is
chosen such that:\footnote{In principle, one should use a one-sided constraint to guarantee feasibility, i.e.: 
$0 <I(X_i;X_{\pi^{(i)}(1)}) - I(T_2^{(i)};X_i)< 0.1\,I(X_i;X_{\pi^{(i)}(1)}).$}
\begin{equation}
\bigl|
I(T_2^{(i)};X_i)
-
I(X_i;X_{\pi^{(i)}(1)})
\bigr|
<
0.1\,I(X_i;X_{\pi^{(i)}(1)}).    
\end{equation}
The approximate third-order common information is then defined as:
\begin{align}\notag
\label{eq:R3approx}
R_3(\mathbf X)
\approx
&\min_{i\in\{1,2,3\}}
\max_{\pi^{(i)}\in S_2^{(i)}}
\min
\Bigl\{I(T_2^{(i)};X_{i}), \\
&\quad
I(T_2^{(i)};X_{\pi^{(i)}(1)}),
I(T_2^{(i)};X_{\pi^{(i)}(2)})
\Bigr\}.
\end{align}

It is straightforward to extend this approach to higher order common information for $n>3$. For example, for $n=4$, one would construct $T^{(i)}_3 = T_2^{(i)} + N$, and perform the information matching with respect to a fourth variable $X_4$. 

\subsection{Obtaining Neural Speech Tracking Responses}
An indication of the strength of speech tracking performance can be obtained by computing the correlations between the electrode signals and the envelopes of the acoustic signals. 
Specifically, let $\rho_{s,t,\tau}(\mathrm{E},\mathrm{env})$ denote the absolute Pearson correlation coefficient between the EEG signal recorded at electrode $\mathrm{E}$ and the corresponding speech envelope for subject $s$, trial $t$, and neural delay $\tau$. Previous studies have shown that cortical tracking of continuous speech is characterized by stimulus-response latencies extending from approximately 100 to 400 ms \cite{LalorFoxe2010}, with particularly strong attention-related effects observed around 200 ms \cite{OSullivan2015}. 

For each subject and trial, absolute correlation values were computed for the four electrode-envelope pairs and these define the speech tracking related response variables:
\begin{align}\notag
&\rho_{s,t,\tau}(\mathrm{FT}_7,\mathrm{env}_T), 
\rho_{s,t,\tau}(\mathrm{FT}_7,\mathrm{env}_D), 
\rho_{s,t,\tau}(\mathrm{T}_7,\mathrm{env}_T),  \\ \label{eq:Dres}
&\rho_{s,t,\tau}(\mathrm{T}_7,\mathrm{env}_D). 
\end{align}
We define the target and distractor speech tracking response variables as follows:
\begin{align}
    D^T(s,t,\tau) &= \frac{1}{2}(\rho_{s,t,\tau}(\mathrm{FT}_7,\mathrm{env}_T) + \rho_{s,t}(\mathrm{T}_7,\mathrm{env}_T)), \\
    D^D(s,t,\tau) &= \frac{1}{2}(\rho_{s,t,\tau}(\mathrm{FT}_7,\mathrm{env}_D) + \rho_{s,t}(\mathrm{T}_7,\mathrm{env}_D)).
\end{align}

\subsection{Statistical Tests}
To investigate whether pairwise and higher-order common information are related to neural speech tracking, we analyzed the trial-to-trial relationship between the information-theoretic quantities $(R_2^T,R_2^D,R_3^T,R_3^D)$ and the corresponding speech-tracking responses 
For each delay, we constructed linear regression models in which the trial-wise speech-tracking responses served as dependent variables and the common-information measures served as predictors. Separate models were fitted for target and distractor responses. To quantify the contribution of pairwise interactions, we first considered models containing only $R_2$. We then considered models containing only $R_3$, as well as combined models containing both $R_2$ and $R_3$. Finally, to determine whether higher-order common information explained variance beyond that captured by pairwise interactions, we evaluated the unique contribution of $R_3$ after controlling for $R_2$ using partial regression analyses.
Specifically, for each subject $s$ and delay $\tau$, we fitted the following models:
\begin{align}
D^T(s,t,\tau) &= \beta_0 + \beta_1 R_2^T(s,t) + \varepsilon_t, \\
D^D(s,t,\tau) &= \beta_0 + \beta_1 R_2^D(s,t) + \varepsilon_t,
\end{align}
for the pairwise-information analysis, and:
\begin{align}
D^T(s,t,\tau) &= \beta_0 + \beta_1 R_3^T(s,t) + \varepsilon_t, \\
D^D(s,t,\tau) &= \beta_0 + \beta_1 R_3^D(s,t) + \varepsilon_t,
\end{align}
for the higher-order analysis. To assess the unique contribution of higher-order common information, we further considered models of the form:
\begin{align}
D^T(s,t,\tau)
&=
\beta_0
+
\beta_1 R_2^T(s,t)
+
\beta_2 R_3^T(s,t)
+
\varepsilon_t, \\
D^D(s,t,\tau)
&=
\beta_0
+
\beta_1 R_2^D(s,t)
+
\beta_2 R_3^D(s,t)
+
\varepsilon_t.
\end{align}
The regression coefficients and coefficients of determination ($\mathrm{R}^2$) were computed separately for each subject and delay. Group-level significance was assessed using cluster-based permutation tests across delays, thereby correcting for multiple comparisons across the temporal dimension. In addition to testing whether the regression models explained significant variance in the target and distractor responses separately, we directly compared target and distractor effects and evaluated whether the unique contribution of $R_3$ remained significant after controlling for $R_2$. The results of the statistical tests are reported in Table~\ref{tab:r2_r3_summary}.

The cluster-based permutation tests revealed significant representations of both pairwise ($R_2$) and higher-order ($R_3$) information throughout the delay interval (100--400 ms). Moreover, $R_3$ explained significant variance beyond that accounted for by $R_2$, indicating the presence of higher-order information that could not be reduced to pairwise interactions alone. On the other hand, no significant target--distractor differences were observed for $R_2$, $R_3$, or Partial $R_3|R_2$ (all cluster-corrected $p > 0.05$), which suggests that the representations of pairwise and higher-order information were comparable for target and distractor stimuli.

\begin{table}[t]
\centering
\caption{Summary of cluster-based permutation test results for 10000 permutations. Significant clusters are reported after cluster-level correction for multiple comparisons across delays.}
\label{tab:r2_r3_summary}
\begin{tabular}{ll}
\hline
Analysis &Cluster $p$ \\
\hline
$R_2$ target & 0.0001 \\
$R_2$ distractor &  0.0006 \\
$R_2$ target -- distractor &  n.s. \\
\hline
$R_3$ target &  0.0001 \\
$R_3$ distractor & 0.0001 \\
$R_3$ target -- distractor & n.s. \\
\hline
Partial $R_3|R_2$ target &  0.0004 \\
Partial $R_3|R_2$ distractor &  0.0010 \\
Partial $R_3|R_2$ target -- distractor & n.s. \\
\hline
\end{tabular}
\end{table}

\appendix

\subsection{Proof of Theorem~\ref{thm:general_gaussian}}

Before proving the theorem, we first need a result that shows that a Gaussian test channel is stage-wise optimal. 
A closely related problem was studied in \cite[Theorem 3]{ErkipCover1998}, who likewise showed that Gaussian channels are worst-case.
For completeness and clarity, we provide a self-contained proof for our problem. Specifically, we show that the IM inequality
achieves equality by every stage-wise optimizer and that every
target-minimal optimizer is equivalent to a scalar Gaussian auxiliary. Thus, if a non-Gaussian stage-wise optimizer exists, it can be replaced by Gaussian one.

\begin{lemma}[Gaussian one-stage extremality]
\label{lem:Gaussian_one_stage}
Let $V,X_1,\ldots,X_m$ be jointly Gaussian scalar random variables,
and suppose $S-V-(X_1,\ldots,X_m)$.
Fix $a\in\{1,\ldots,m\}$, and write $r_a\triangleq |\operatorname{corr}(V,X_a)|$.
Then, for every $S$ satisfying $I(S;V)\leq\Delta$,
we have
\begin{equation}
I(S;X_a)
\leq
-\frac12\log_2\!\left(
1-r_a^2\left(1-2^{-2\Delta}\right)
\right).
\label{eq:Gaussian_one_stage_bound}
\end{equation}
Equality is attained by a Gaussian test channel:
\begin{equation}
G=\alpha V+Z,
\label{eq:Gaussian_test_channel}
\end{equation}
where $Z$ is zero-mean Gaussian and independent of
$(V,X_1,\ldots,X_m)$, with its variance chosen such that $I(G;V)=\Delta$.
Moreover, suppose that $0<r_a<1, 0<\Delta<\infty$. If equality holds in \eqref{eq:Gaussian_one_stage_bound}, then $I(S;V)=\Delta$
and there exists a scalar Gaussian statistic $G=g(S)$ such that:
\begin{align}
&I(G;V)=I(S;V),                                      \label{eq:G_sufficient_V}\\
&I(G;X_j)=I(S;X_j),\quad j=1,\ldots,m,              \label{eq:G_sufficient_sources}\\
&(X_1,\ldots,X_m)-G-S.                               \label{eq:G_sufficient_MC}
\end{align}
The above implies that if $S\in\mathsf M(X_a)$, then $S$ and $G$ are
equivalent up to almost-surely invertible transformations.
\end{lemma}

\begin{IEEEproof}
By applying invertible affine transformations, we may by Lemma~\ref{lem:invertible_invariance} assume that
$V$ and $X_a$ have zero mean and unit variance. After possibly
replacing $V$ by $-V$, we may write $X_a=r_aV+\sqrt{1-r_a^2}\,W$, 
where $W\sim\mathcal N(0,1)$ is independent of $V$. Since $S-V-X_a$, $W$ may be chosen independent of $(V,S)$.
The conditional entropy-power inequality gives \cite{Costa1985EPI}:
\begin{equation}
N(X_a|S)
\geq
r_a^2N(V|S)+(1-r_a^2)N(W),
\end{equation}
where $N(V|S) \triangleq \frac{1}{2\pi e}2^{2h(V|S)}$.
Since $V\sim\mathcal N(0,1)$, and by defining $\delta\triangleq I(S;V)\leq\Delta$, we have that:
\begin{equation}
N(V|S)
=
\frac{1}{2\pi e}2^{2[h(V)-I(S;V)]}
=
2^{-2\delta},
\end{equation}
and $N(W)=1$. Hence, we can write:
\begin{equation}
N(X_a|S)
\geq
1-r_a^2+r_a^2 2^{-2\delta}.
\end{equation}
It follows that:
\begin{align}
I(S;X_a)
&=h(X_a)-h(X_a|S)\\
&\leq
-\frac12\log_2\!\left(
1-r_a^2+r_a^2 2^{-2\delta}
\right).
\label{eq:Gaussian_bound_delta}
\end{align}
For $r_a>0$, the right-hand side of
\eqref{eq:Gaussian_bound_delta} is strictly increasing in $\delta$.
Since $\delta\leq\Delta$,
\begin{equation}
I(S;X_a)
\leq
-\frac12\log_2\!\left(
1-r_a^2+r_a^2 2^{-2\Delta}
\right),
\end{equation}
which proves \eqref{eq:Gaussian_one_stage_bound}.

Now let $G=\alpha V+Z$, where $Z$ is Gaussian and independent of the complete source vector,
with its variance chosen so that $I(G;V)=\Delta$.
Then $(V,G)$ is jointly Gaussian, and $V|G=g$ is Gaussian with a
variance that does not depend on $g$. Furthermore, $W$ is independent
of $(V,G)$. Therefore, conditioned on $G=g$, $r_aV \perp \sqrt{1-r_a^2}\,W$ 
are independent Gaussian random variables. Equality therefore holds
in the conditional EPI, and $G$ achieves the upper bound in 
\eqref{eq:Gaussian_one_stage_bound}.

Suppose now that $0<r_a<1$ and equality holds in
\eqref{eq:Gaussian_one_stage_bound}. Since the bound in
\eqref{eq:Gaussian_bound_delta} is strictly increasing in $\delta$,
equality requires $I(S;V)=\delta=\Delta$.
It also requires equality in the conditional EPI, which further imply that, for almost every $s$,
\begin{equation}
V|S=s\sim\mathcal N(m(s),\sigma_E^2),
\end{equation}
where the conditional variance $\sigma_E^2$ does not depend on $s$.
Define $G\triangleq\mathbb E[V|S]$.
Then, $V=G+E$,
where $E\sim\mathcal N(0,\sigma_E^2)$
is independent of $S$. Since $G$ is a function of $S$, $E$ is
also independent of $G$. Because $V$ and $E$ are Gaussian,
Cramér's theorem implies that $G$ is Gaussian \cite{cramer1970random}.

The conditional distribution of $V$ given $S=s$ depends on $s$ only
through $G=g(s)$. Therefore, $V-G-S$. Combining this with $S-V-(X_1,\ldots,X_m)$ gives $(X_1,\ldots,X_m)-G-S$.
Since $G=g(S)$, we therefore have that
\begin{align*}
I(S;V)
&=I(G;V)+I(S;V|G)
=I(G;V),\\
I(S;X_j)
&=I(G;X_j)+I(S;X_j|G)
=I(G;X_j),
\end{align*}
for $j=1,\ldots,m$.
Thus, every equality-achieving feasible auxiliary $S$ contains a
Gaussian source-sufficient statistic $G=g(S)$. If, in addition,
$S\in\mathsf M(X_a)$, then $X_a-G-S$ and target minimality imply
that $S=f(G)$ almost surely. Hence $S$ and $G$ are equivalent up
to almost-surely invertible transformations.

\end{IEEEproof}

\begin{IEEEproof}[Proof of Theorem~\ref{thm:general_gaussian}]
By Lemma~\ref{lem:invertible_invariance}, the HCI is invariant under
invertible transformations of the individual source variables.
Therefore, we may normalize the variables so that $\mathbb E[X_j^2]=1, j=1,\ldots,n.$
Fix a reference variable $X_i$ and an ordering $\pi^{(i)}(1),\ldots,\pi^{(i)}(n-1)$
of the remaining source variables, and set $T_1^{(i)}=X_i$.
At stage $\ell$, let $a_\ell\triangleq\pi^{(i)}(\ell)$ and $\Delta_\ell \triangleq I(T_\ell^{(i)};X_{a_\ell})$.

Using induction, we now suppose that $T_\ell^{(i)}$ is jointly Gaussian with
$\mathbf X$. For every feasible $S$, define $\delta\triangleq I(S;T_\ell^{(i)})\leq\Delta_\ell$.
Applying Lemma~\ref{lem:Gaussian_one_stage} using $V=T_\ell^{(i)}$
shows that:
\begin{equation}
I(S;X_{a_\ell})
\leq
-\frac12\log_2\!\left(
1-r_\ell^2+r_\ell^2 2^{-2\Delta_\ell}
\right),
\end{equation}
where $r_\ell \triangleq | \operatorname{corr} (T_\ell^{(i)},X_{a_\ell}) |$.
The upper bound is attained by a Gaussian test channel satisfying $I(T_{\ell+1}^{(i)};T_\ell^{(i)}) = \Delta_\ell$.
Thus, the IM inequality achieves its maximum by a Gaussian optimizer.
Moreover, in the nondegenerate case, every target-minimal stage-wise
optimizer must attain this equality and by Lemma~\ref{lem:Gaussian_one_stage}, the optimizer is almost surely equivalent to a
scalar Gaussian statistic. Starting from $T_1^{(i)}=X_i$, where $X_i$ is Gaussian, and applying the above induction step,  shows that every target-minimal stage-wise optimizer may be
represented as a scalar random variable jointly Gaussian with
$\mathbf X$.

We now proceed finding a closed-form expression for the HCI. For each fixed $i$, define $r_{i,j}\triangleq|\rho_{i,j}|, j\neq i$, and let
$ r_{i,(1)} \leq r_{i,(2)} \leq\cdots\leq r_{i,(n-1)}$ denote the order statistics of $\{r_{i,j}:j\neq i\}$. Also define $q_\ell^{(i)} \triangleq | \operatorname{corr}(T_\ell^{(i)},X_i)|$.
Since $T_1^{(i)}=X_i$, it follows that $q_1^{(i)}=1$ for any $i$.

The jointly Gaussian Markov chain $T_\ell^{(i)}-X_i-X_{a_\ell}$ implies that the correlation between $T_\ell^{(i)}$ and $X_{a_\ell}$ can be written as a product of the correlations between $(T_\ell^{(i)},X_i)$ and $(X_i,X_{a_\ell})$ that is:
\begin{equation}
\left|
\operatorname{corr}
(T_\ell^{(i)},X_{a_\ell})
\right|
=
q_\ell^{(i)}r_{i,a_\ell}.
\end{equation}
In the jointly Gaussian case, the mutual information only depends upon the correlation between the variables. We therefore obtain:
\begin{equation}
\Delta_\ell
=
I(T_\ell^{(i)};X_{a_\ell})
=
-\frac12\log_2\!\left(
1-\left(q_\ell^{(i)}r_{i,a_\ell}\right)^2
\right).
\label{eq:stage_matching_Gaussian}
\end{equation}

As shown by Lemma~\ref{lem:Gaussian_one_stage}, every stage-wise optimizer achieves the maximum in the IM inequality, i.e., $I(T_{\ell+1}^{(i)};T_\ell^{(i)})
= \Delta_\ell$.
Since the variables are jointly Gaussian, this is equivalent to: $|\operatorname{corr} (T_{\ell+1}^{(i)},T_\ell^{(i)})|
= q_\ell^{(i)}r_{i,a_\ell}$.
Using the jointly Gaussian Markov chain $T_{\ell+1}^{(i)}-T_\ell^{(i)}-X_i$,
we obtain the following recursive form for the correlation:
\begin{align}
q_{\ell+1}^{(i)}
&=
\left|
\operatorname{corr}
(T_{\ell+1}^{(i)},X_i)
\right|\\
&=
\left|
\operatorname{corr}
(T_{\ell+1}^{(i)},T_\ell^{(i)})
\right|
\left|
\operatorname{corr}
(T_\ell^{(i)},X_i)
\right|\\
&=
\left(q_\ell^{(i)}\right)^2r_{i,a_\ell}.
\label{eq:Gaussian_q_recursion}
\end{align}
Iterating this recursion over the $n-2$ processed variables yields:
\begin{equation}
q_{n-1}^{(i)}
=
\prod_{\ell=1}^{n-2}
r_{i,a_\ell}^{\,2^{n-2-\ell}}.
\label{eq:terminal_q}
\end{equation}

For every $(j,i)$, the jointly Gaussian Markov chain $T_{n-1}^{(i)}-X_i-X_j$ leads to:
\begin{equation}
\left|
\operatorname{corr}
(T_{n-1}^{(i)},X_j)
\right|
=
q_{n-1}^{(i)}r_{i,j},
\end{equation}
where for $j=i$, we have $r_{i,i}=1$, and the absolute correlation simplifies to $q_{n-1}^{(i)}$.
Since $r_{i,j}\leq1$ and by using the order statistics, the smallest terminal correlation is $q_{n-1}^{(i)}r_{i,(1)}$, which implies that:
\begin{align}
\min_{1\leq j\leq n}
I(T_{n-1}^{(i)};X_j)
&=
-\frac12\log_2\!\left(
1-r_{i,(1)}^2
\left(q_{n-1}^{(i)}\right)^2
\right)\\
&=
-\frac12\log_2\!\left(
1-r_{i,(1)}^2
\prod_{\ell=1}^{n-2}
r_{i,a_\ell}^{\,2^{n-1-\ell}}
\right).
\label{eq:fixed_order_value}
\end{align}

It remains to optimize over the ordering of the correlations within the product. These correlations must be processed in decreasing order since for $0\leq a\leq b\leq 1, u\leq v$, we have $a^u b^v\geq a^v b^u$.
Thus, the largest correlation receives the largest exponent, and an
optimal processing order is $r_{i,(n-1)},r_{i,(n-2)},\ldots,r_{i,(2)}$.
Substituting this ordering into \eqref{eq:fixed_order_value} gives:
\begin{equation}
\max_{\pi^{(i)}}
\min_j I(T_{n-1}^{(i)};X_j)
=
-\frac12\log_2\!\left(
1-r_{i,(1)}^2
\prod_{m=2}^{n-1}
r_{i,(m)}^{\,2^{m-1}}
\right).
\end{equation}
Finally, minimizing over the reference variable $i$ yields
\begin{equation}
R(X_1,\ldots,X_n)
=
\min_{1\leq i\leq n}
-\frac12\log_2\!\left(
1-r_{i,(1)}^2
\prod_{m=2}^{n-1}
r_{i,(m)}^{\,2^{m-1}}
\right).
\end{equation}
If some $\rho_{ij}=0$, the pairwise upper bound gives $R=0$. If some but not all source variables are maximally correlated, 
their corresponding absolute correlations are equal to one and the corresponding HCI is finite. If $\rho_{i,j}=1$ for all $(i,j)$, then the HCI is unbounded, i.e., $R(X_1,\dotsc, X_n) = \infty$.
\end{IEEEproof}

\subsection{Proof of Theorem~\ref{thm:Gaussian_n3}}

Fix a reference index $i\in\{1,2,3\}$. Let $a\neq i$ denote the
source used in the single IM constraint, and let $b$ be the remaining
index, so that $\{i,a,b\}=\{1,2,3\}$.
For this fixed reference and ordering, define the common feasible set:
\begin{equation}
\mathcal F_{i,a}
\triangleq
\left\{
S:
\begin{array}{l}
S-X_i-(X_1,X_2,X_3),\\[0.2em]
I(S;X_i)\leq I(X_i;X_a)
\end{array}
\right\}.
\label{eq:Gaussian_common_feasible_set}
\end{equation}

Both the upper bound and the HCI optimize over $\mathcal F_{i,a}$, but they
select auxiliaries differently. The fixed-reference, fixed-order upper bound 
value is:
\begin{equation}
V_{i,a}^{\mathrm{ub}}
\triangleq
\sup_{S\in\mathcal F_{i,a}}
\min_{j\in\{1,2,3\}}I(S;X_j).
\label{eq:Gaussian_global_value}
\end{equation}
On the other hand, for the HCI, we first define the set of stage-wise optimizers:
\begin{equation}
\mathcal A_{i,a}
\triangleq
\arg\sup_{S\in\mathcal F_{i,a}} I(S;X_a).
\label{eq:Gaussian_stagewise_set}
\end{equation}
The corresponding fixed-reference, fixed-order HCI value is then:
\begin{equation}
V_{i,a}^{\mathrm{HCI}}
\triangleq
\sup_{S\in\mathcal A_{i,a}}
\min_{j\in\{1,2,3\}}I(S;X_j).
\label{eq:Gaussian_SW_value}
\end{equation}
Because $\mathcal A_{i,a}\subseteq\mathcal F_{i,a}$,
we immediately have:
\begin{equation}
V_{i,a}^{\mathrm{HCI}}\leq V_{i,a}^{\mathrm{ub}}.
\label{eq:fixed_HCI_less_global}
\end{equation}

We next show that equality holds in
\eqref{eq:fixed_HCI_less_global}. Define $\Delta_a\triangleq I(X_i;X_a)$
and:
\begin{equation}
\Phi(r,\Delta)
\triangleq
-\frac12\log_2
\left[
1-r^2\left(1-2^{-2\Delta}\right)
\right],
\quad r\in[0,1].
\label{eq:Gaussian_Phi}
\end{equation}
For the moment, suppose that $0<|\rho_{ia}|<1$, so that
$0<\Delta_a<\infty$.
Let $S$ be an element of $\mathcal F_{i,a}$, and set $\Delta_S\triangleq I(S;X_i)$.
Notice that $S$ is not assumed to optimize either
\eqref{eq:Gaussian_global_value} or
\eqref{eq:Gaussian_stagewise_set}. Since $S$ is feasible, $\Delta_S\leq\Delta_a$.
Moreover, for every $j\neq i$, the Markov chain $S-X_i-X_j$
and Lemma~\ref{lem:Gaussian_one_stage} give:
\begin{equation}
I(S;X_j)
\leq
\Phi\!\left(|\rho_{ij}|,\Delta_S\right).
\label{eq:Gaussian_one_stage_each_source}
\end{equation}
For $j=i$, the expression is still valid after setting $|\rho_{ii}|=1$, because $\Phi(1,\Delta_S)=\Delta_S=I(S;X_i)$.
Since $\Phi(r,\Delta)$ is nondecreasing in $\Delta$, we obtain:
\begin{equation}
I(S;X_j)
\leq
\Phi\!\left(|\rho_{ij}|,\Delta_a\right),
\quad j=1,2,3.
\label{eq:Gaussian_common_upper_bound}
\end{equation}
Therefore, every $S\in\mathcal F_{i,a}$ satisfies:
\begin{equation}
\min_j I(S;X_j)
\leq
\min_j
\Phi\!\left(|\rho_{ij}|,\Delta_a\right).
\label{eq:Gaussian_global_upper_bound}
\end{equation}
It follows that:
\begin{equation}
V_{i,a}^{\mathrm{ub}}
\leq
\min_j
\Phi\!\left(|\rho_{ij}|,\Delta_a\right).
\label{eq:Gaussian_V_upper}
\end{equation}

At this point, we construct a Gaussian auxiliary:
\begin{equation}
G_{i,a}=\gamma X_i+Z,
\label{eq:Gaussian_test_channel_fixed}
\end{equation}
where $Z$ is zero-mean Gaussian and independent of
$(X_1,X_2,X_3)$, with its variance chosen such that:
\begin{equation}
I(G_{i,a};X_i)=\Delta_a.
\label{eq:Gaussian_test_rate}
\end{equation}
Thus, $G_{i,a}\in\mathcal F_{i,a}$ since both constraints in \eqref{eq:Gaussian_common_feasible_set} are satisfied. The equality statement in
Lemma~\ref{lem:Gaussian_one_stage} gives, for every $j$:
\begin{equation}
I(G_{i,a};X_j)
=
\Phi\!\left(|\rho_{ij}|,\Delta_a\right),
\quad j=1,2,3.
\label{eq:Gaussian_simultaneous_attainment}
\end{equation}
It therefore follows that:
\begin{equation}
\min_j I(G_{i,a};X_j)
=
\min_j
\Phi\!\left(|\rho_{ij}|,\Delta_a\right).
\label{eq:Gaussian_global_attainment}
\end{equation}
Comparing \eqref{eq:Gaussian_V_upper} and
\eqref{eq:Gaussian_global_attainment} shows that $G_{i,a}$ attains
the global value:
\begin{equation}
V_{i,a}^{\mathrm{ub}}
=
\min_j
\Phi\!\left(|\rho_{ij}|,\Delta_a\right).
\label{eq:Gaussian_global_exact}
\end{equation}

It remains to verify that the same auxiliary is stage-wise optimal.
For any $S\in\mathcal F_{i,a}$,
\begin{align}
I(S;X_a)
&\leq
\Phi\!\left(|\rho_{ia}|,\Delta_S\right)\\
&\leq
\Phi\!\left(|\rho_{ia}|,\Delta_a\right)\\
&=
I(G_{i,a};X_a).
\end{align}
Hence $G_{i,a}\in\mathcal A_{i,a}$.
The globally optimal auxiliary $G_{i,a}$ is therefore also admissible
in the HCI optimization. It follows that:
\begin{equation}
V_{i,a}^{\mathrm{HCI}}
\geq
\min_j I(G_{i,a};X_j)
=
V_{i,a}^{\mathrm{ub}}.
\end{equation}
Together with \eqref{eq:fixed_HCI_less_global}, this gives:
\begin{equation}
V_{i,a}^{\mathrm{HCI}}=V_{i,a}^{\mathrm{ub}}
\label{eq:fixed_problem_equality}
\end{equation}
for every fixed reference $i$ and ordering $(i,a,b)$.
For $n=3$, maximizing over orderings is equivalent to maximizing over
the two possible choices of $a\neq i$. Therefore,
\begin{align}
R_{\mathrm{ub}}(X_1,X_2,X_3)
&=
\min_i\max_{a\neq i}V_{i,a}^{\mathrm{ub}},\\
R(X_1,X_2,X_3)
&=
\min_i\max_{a\neq i}V_{i,a}^{\mathrm{HCI}}.
\end{align}
Applying \eqref{eq:fixed_problem_equality} for every
$(i,a)$ proves:
\begin{equation}
R(X_1,X_2,X_3)
=
R_{\mathrm{ub}}(X_1,X_2,X_3).
\end{equation}

The boundary cases are obtained directly. If $|\rho_{ia}|=0$, then
$\Delta_a=0$, and every feasible auxiliary is independent of $X_i$
and hence of the complete source vector. If $|\rho_{ia}|=1$, then
$X_i$ and $X_a$ are almost surely invertibly related after
normalization, and the identity auxiliary $G_{i,a}=X_i$ is both
globally and stage-wise optimal. Thus,
\eqref{eq:fixed_problem_equality} also holds in the boundary cases.

\hfill$\square$

\subsection{Proof of Theorem~\ref{thm:Bernoulli_HCI}}
Similarly to the Gaussian case, we will first establish a one-stage extremal result, which proves that the binary test channel is stage-wise optimal. 
We will prove this using Mrs. Gerber's Lemma (MGL) \cite{WynerZiv1973MGL}. 

\begin{lemma}[Binary one-stage extremality]
\label{lem:binary_one_stage}
Let $V\sim\mathrm{Bern}(\tfrac12)$, and let $X=V\oplus N, N\sim\mathrm{Bern}(p), 0\leq p\leq\tfrac12$, where $N\perp V$. Let $S$ satisfy $S-V-X$ and $I(S;V)=\Delta$. Then
\begin{equation}
I(S;X)
\leq
1-h_2\!\left(
p\star h_2^{-1}(1-\Delta)
\right),
\label{eq:binary_one_stage_bound}
\end{equation}
where equality is attained by the binary symmetric test channel $G=V\oplus E, E\sim\mathrm{Bern}(\varepsilon), \varepsilon=h_2^{-1}(1-\Delta),$ and $E$ is independent of $(V,X)$.
Moreover, suppose that $S-V-\mathbf X, \mathbf X=(X_1,\ldots,X_n)$, and that one of the components of $\mathbf X$ is $X$. 
If $0<p<\frac12, 0<\Delta<1,$ and equality holds in \eqref{eq:binary_one_stage_bound}, then there
exists a binary statistic $G=g(S)$ such that:
\begin{align}
&I(G;V)=I(S;V),\\
&I(G;X_j)=I(S;X_j),
\quad j=1,\ldots,n,\\
&\mathbf X-G-S.
\end{align}
It follows that if $S$ is target-minimal, then $S$ and $G$ are equivalent up to an almost-surely invertible transformation.
\end{lemma}

\begin{IEEEproof}
Since $V\sim\mathrm{Bern}(\tfrac12)$ and $I(S;V)=\Delta$, it follows that $H(V|S)=1-\Delta$.
Moreover, $X=V\oplus N$, where
$N\sim\mathrm{Bern}(p)$ is independent of $(V,S)$. The conditional
form of Mrs.\ Gerber's lemma therefore gives \cite{WynerZiv1973MGL}:
\begin{equation}\label{eq:cmrsG}
H(X|S)
\geq
h_2\!\left(
p\star h_2^{-1}(H(V|S))
\right)
=
h_2\!\left(
p\star h_2^{-1}(1-\Delta)
\right).    
\end{equation}
Since $X\sim\mathrm{Bern}(\tfrac12)$, $H(X)=1$, and hence:
\begin{equation}
I(S;X)
=
1-H(X|S)
\leq
1-h_2\!\left(
p\star h_2^{-1}(1-\Delta)
\right).
\end{equation}

To establish achievability, let $G=V\oplus E, E\sim\mathrm{Bern}(\varepsilon), \varepsilon=h_2^{-1}(1-\Delta)$, where $E$ is independent of $(V,N)$. Then
\begin{equation}
I(G;V)=1-h_2(\varepsilon)=\Delta.
\end{equation}
Furthermore, $G\oplus X=E\oplus N\sim
\mathrm{Bern}(p\star\varepsilon)$, and therefore:
\begin{equation}
I(G;X)
=
1-h_2(p\star\varepsilon)
=
1-h_2\!\left(
p\star h_2^{-1}(1-\Delta)
\right),
\end{equation}
which shows that $G$  achieves the upper bound in \eqref{eq:binary_one_stage_bound}.
Now suppose that $0<p<\tfrac12, 0<\Delta<1$,
and that equality holds in
\eqref{eq:binary_one_stage_bound}. 
Then, we must have that $P(V=1|S)\in\{\varepsilon,1-\varepsilon\}$. Thus, we can define: 
\begin{equation}
G=g(S)
\triangleq
\begin{cases}
0,&P(V=1|S)=\varepsilon,\\
1,&P(V=1|S)=1-\varepsilon.
\end{cases}
\end{equation}
The conditional distribution of $V$ given $S=s$ is therefore identical to the conditional distribution of $V$ given $G$, i.e., $P_{V|S=s}=P_{V|G=g(s)}$, which implies the following Markov chain: $V-G-S$. Moreover, 
since $G$ is a deterministic function of $S$, it follows that $I(G;V)=I(S;V)$.
Furthermore, $S-V-\mathbf X$ together with $V-G-S$ implies $\mathbf X-G-S$.
Since $G$ is also a function of $S$, we obtain $I(G;X_j)=I(S;X_j), j=1,\ldots,n$.
Thus every auxiliary $S$ that achieves equality in the binary one-stage
bound can be replaced by the binary statistic $G$ without changing
its mutual information with $V$ or any source component.
\end{IEEEproof}

\begin{IEEEproof}[Proof of Theorem~\ref{thm:Bernoulli_HCI}]
We first fix a reference index $i$ and an ordering $\pi^{(i)}(1),\ldots,\pi^{(i)}(n-1)$
of the remaining sources. Set $T_1^{(i)}=X_i$, and write $a_\ell\triangleq\pi^{(i)}(\ell), \ell=1,\ldots,n-2$.
From the theorem, we know that $X_j=U\oplus N_j, c_j=1-2p_j$.
For every $j\neq i$, it can be observed that the pair $(X_i,X_j)$ is related by a BSC with
bias $1-2(p_i\star p_j)=c_ic_j$.

By Lemma~\ref{lem:binary_one_stage}, the stage-wise optimum under the
IM constraint is achieved by a BSC test channel, and every stage-wise optimizer is equivalent to a uniform binary auxiliary.
Thus, the auxiliaries may be represented as a cascade of BSCs.
Define the bias between $T_\ell^{(i)}$ and $X_i$ by $q_\ell^{(i)}
\triangleq
1-2P(T_\ell^{(i)}\neq X_i)$.
Since $T_1^{(i)}=X_i$, $q_1^{(i)}=1$.
The BSC bias between $T_\ell^{(i)}$ and the current target
$X_{a_\ell}$ is $q_\ell^{(i)}c_ic_{a_\ell}$. Thus, $I(T_\ell^{(i)};X_{a_\ell}) = 1-h_2(\frac{1-q_\ell^{(i)}c_ic_{a_\ell}}{2})$.

The one-stage lemma shows that the optimal test channel from
$T_\ell^{(i)}$ to $T_{\ell+1}^{(i)}$ is a BSC having the same bias
$q_\ell^{(i)}c_ic_{a_\ell}$. Since BSC biases multiply under
cascading, it follows that:
\begin{equation}
q_{\ell+1}^{(i)}
=
\bigl(q_\ell^{(i)}\bigr)^2c_ic_{a_\ell}.
\label{eq:binary_bias_recursion}
\end{equation}
Iterating this recursion yields:
\begin{equation}
q_{n-1}^{(i)}
=
c_i^{\,2^{n-2}-1}
\prod_{\ell=1}^{n-2}
c_{a_\ell}^{\,2^{n-2-\ell}}.
\label{eq:binary_terminal_bias}
\end{equation}
For $j\neq i$, the terminal bias between $T_{n-1}^{(i)}$ and
$X_j$ is $q_{n-1}^{(i)}c_ic_j$,
whereas its bias with $X_i$ is $q_{n-1}^{(i)}$. Hence, by using order statistics, the smallest
terminal bias is:
\begin{equation}
q_{n-1}^{(i)}c_i\min_{j\neq i}c_j
=
q_{n-1}^{(i)}c_i d_{i,(1)}.
\end{equation}
Since the mutual information of two uniform binary variables is
increasing in their BSC bias, we obtain:
\begin{equation}
\min_j I(T_{n-1}^{(i)};X_j)  = 1-h_2\!\left(
\frac{1-\beta_i(\pi^{(i)})}{2}
\right),
\end{equation}
where
\begin{equation}
\beta_i(\pi^{(i)})
=
d_{i,(1)}
c_i^{\,2^{n-2}}
\prod_{\ell=1}^{n-2}
c_{a_\ell}^{\,2^{n-2-\ell}}.
\label{eq:fixed_order_binary_bias}
\end{equation}

It remains to optimize the ordering of biases within the product. The product in
\eqref{eq:fixed_order_binary_bias} is maximized by assigning the largest exponent to
the largest remaining bias. Thus, the optimal processing order is $d_{i,(n-1)},d_{i,(n-2)},\ldots,d_{i,(2)}$.
Substitution into \eqref{eq:fixed_order_binary_bias} gives:
\begin{equation}
\beta_i
=
d_{i,(1)}
c_i^{\,2^{n-2}}
\prod_{m=2}^{n-1}
d_{i,(m)}^{\,2^{m-2}}.
\end{equation}
Therefore, for reference $X_i$, we have that:
\begin{equation}
\max_{\pi^{(i)}}
\sup_{T\in\mathcal T_i^{\pi^{(i)}}}
\min_j I(T;X_j)
=
1-h_2\!\left(\frac{1-\beta_i}{2}\right).
\end{equation}
Finally, minimizing over the reference index yields:
\begin{equation}
R(X_1,\ldots,X_n)
=
\min_{1\leq i\leq n}
\left[
1-h_2\!\left(\frac{1-\beta_i}{2}\right)
\right].
\end{equation}
The boundary cases $c_j\in\{0,1\}$ follow directly from the same
recursion by continuity arguments. 
\end{IEEEproof}

\section*{Conclusions}
We introduced the concept of higher-order common information for quantifying shared information among $n$ random variables. Our results enable the assessment of multivariate dependencies beyond pairwise interactions. Additionally, we proposed a practical method for approximating the higher-order common information from arbitrary data, providing a new tool for uncovering inherent structure in complex, real-world datasets.

\section*{Acknowledgment}
The author would like to sincerely thank the referees for their critical and insightful comments on a previous version of this manuscript, which greatly improved the clarity and the quality of the presentation and sharpened the definition of the higher-order common information. To further improve the presentation, fixing language issues, and correcting textual and mathematical typos, the author used ChatGPT-5.6-sol. 
 
The author also made use of ChatGPT-5.6-sol to aid in solving two technical issues in the paper related to: i) monotonicity, and ii) private randomness. While no direct solutions to these issues were found, ChatGPT suggested instead to constrain the optimization problem by introducing two technical requirements. The first technical requirement was universal source dominance introduced in Lemma~\ref{lem:HCI_monotonicity},  which made it possible prove monotonicity in the sense of $R(X_1,\dotsc, X_{n-1})\geq R(X_1,\dotsc, X_{n})$. 
The second technical requirement was target-minimality introduced in \eqref{eq:MX}, which exclude certain solutions containing source-irrelevant information such as those dominated by separable private randomness from the stochastic kernels. 
The author takes full responsibility for the final manuscript.

This work was partially supported by the Novo Nordisk Foundation New Exploratory Research and Discovery Grant, INFOCEPTION: Grant number NNF25OC0100334, and partially supported by the Centre for Acoustic Signal Processing Research (CASPR)
sponsored by the William Demant Foundation Case No. 20-2428.

\bibliographystyle{ieeetr}
\bibliography{sample}

\end{document}